\documentclass[prl,twocolumn,superscriptaddress,showpacs]{revtex4}
\usepackage{graphicx}
\usepackage{epsfig}
\usepackage{amssymb,amsmath,amsfonts,hyperref}
\usepackage{wasysym}
\usepackage{latexsym}
\usepackage{verbatim}
\usepackage[normalem]{ulem}
\usepackage{color}

\newcommand{\be}{\begin{eqnarray}}
\newcommand{\ee}{\end{eqnarray}}

\begin{document}
\title{Columnar order and Ashkin-Teller criticality in mixtures of hard-squares and dimers}
\author{Kabir Ramola}
\affiliation{\small{Laboratoire de Physique Th\'{e}orique et Mod\`{e}les
Statistiques, UMR 8626, Universit\'{e} Paris-Sud 11 and CNRS,
B\^{a}timent 100, Orsay F-91405, France}}
\author{Kedar Damle}
\affiliation{\small{Tata Institute of Fundamental Research, 1 Homi Bhabha Road, Mumbai, India 400005}}
\author{Deepak Dhar}
\affiliation{\small{Tata Institute of Fundamental Research, 1 Homi Bhabha Road, Mumbai, India 400005}}
\begin{abstract}
We show that critical exponents of the transition to columnar order in a
{\em mixture} of $2 \times 1$ dimers and $2 \times 2$ hard-squares on the square lattice 
{\em depends on the composition of the mixture} in exactly the manner predicted by the theory of Ashkin-Teller criticality, including in the hard-square
limit. This result settles the question regarding the nature of the transition in the hard-square lattice gas. It also provides the first example of a polydisperse system
whose critical properties depend on composition. Our ideas also lead to some interesting
predictions  for a class of frustrated quantum magnets that exhibit columnar ordering of the bond-energies at low temperature.
\end{abstract}


\maketitle

{\em Introduction:}  In materials which exhibit a continuous transition from a low-density fluid to an ordered
high-density crystalline state  with spontaneous symmetry breaking, critical properties in the vicinity of the transition are generally independent of microscopic details such as chemical composition and
precise form of the interactions. Indeed, in
the standard theory of such critical phenomena, these properties are generally expected to depend only on the symmetries of the ordered
state. This {\em universality} of critical properties makes it
possible to understand such behaviours in terms of simple models. Lattice-gas models of hard-core particles, with different
sizes and shapes of the excluded-volume region around each particle, provide
many paradigmatic examples of such continuous transitions from a low-density
fluid to a high-density ordered state~\cite{Baxter_hardhexagons,Verberkmoes_Nienhuis, Dickman,Barnes_etal,Ghosh_Dhar,Kundu_etal,Kundu_Rajesh, Nath_Rajesh,Pankov_Moessner_Sondhi}. 

One such simple lattice-gas model, of $2\times2$ hard-squares on the square lattice, has long been of special interest and some controversy. Here, the crystalline state has a sliding instability that leads to long-range columnar (stripe) order in the high-density phase~\cite{Bellemans_Nigam,Ree_Chestnut,Nisbet_Farquhar,Aksenenko_Shulepov,Lafuente_Cuesta,Fernandes_Arenzon_Levin,Zhitomirsky_Tsunetsugu,Feng_Blote_Nienhuis,Ramola_Dhar}. General symmetry arguments~\cite{Domany_Riedel,Domany_Schick_Walker_Griffiths} suggest
that the transition to this columnar ordered phase should provide an example of
``Ashkin-Teller''  (AT) critical behaviour~\cite{Ashkin_Teller,Baxter_PRL,Kadanoff_Wegner,Jose_Kadanoff_Kirkpatrick_Nelson,Kadanoff_PRL,Kadanoff_JPhysA,Kadanoff_Annals,Kadanoff_Brown,Elitzur_Pearson_Shigemitsu,Cardy,Rujan_Williams_Frisch_Forgacs,Kohmoto_denNijs_Kadanoff,Boyanovsky,Delfino,Alet_etal_PRL,Alet_etal_PRB,Papanikolaou_Luijten_Fradkin,Taroni_Bramwell_Holdsworth}.  Such Ashkin-Teller transitions
are interesting exceptions to universality, since the correlation length for columnar
order is expected to grow with a power-law exponent $\nu$ that depends on
microscopic details.
In light of this, it is surprising that several large-scale Monte-Carlo simulations~\cite{Fernandes_Arenzon_Levin,Zhitomirsky_Tsunetsugu,Feng_Blote_Nienhuis} found critical properties that are very close to those of a two-dimensional Ising model. Some of these~\cite{Fernandes_Arenzon_Levin} favoured an Ising critical point, while others
identified small  deviations from Ising behaviour~\cite{Zhitomirsky_Tsunetsugu,Feng_Blote_Nienhuis}. 

\begin{figure}
{\includegraphics[width=\columnwidth]{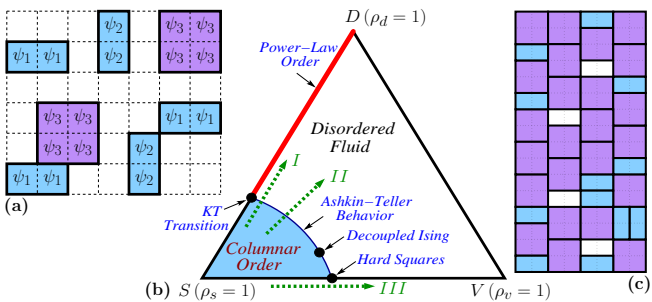}}
\caption{
(Color online) (a) Part of a low-density configuration of $2 \times 1$ tiles (dimers) and $2 \times 2$ tiles (hard-squares) on the square-lattice, also showing values of the columnar
order parameter field $\psi(\vec{r})$ (see Eqn~{\protect{\ref{psidefinition}}}).
(b) Schematic phase diagram of $Z_{dsv}$.  $\rho_s,\rho_d$, and $\rho_v$ are the densities of squares, dimers, and vacancies respectively, with $\rho_s + \rho_d +\rho_v =1$.
Monte-Carlo results along the cuts {\it I}, {\it II} and {\it III} are discussed in text. (c) Columnar ordered high-density configuration, with stripes running
in the vertical direction. }
\label{Fig1}
\end{figure}

In this Letter, we show that critical exponents of the transition to columnar order in a more general {\em mixture} of $2 \times 1$ dimers and $2 \times 2$ hard-squares on the square lattice (Fig~\ref{Fig1} a) {\em depends on the composition of the mixture} in exactly the manner predicted by the theory of Ashkin-Teller criticality, including in the hard-square
limit. This result settles the question regarding the nature of the transition in the hard-square lattice gas. It also provides the first example of a polydisperse system
whose critical properties depend on composition. Our ideas also lead to some interesting
predictions  for a class of frustrated quantum magnets that exhibit columnar ordering of the bond-energies at low temperature.

The original
hard-square lattice-gas corresponds to the boundary-line $VS$ in the
phase-diagram  (Fig~\ref{Fig1} b) of this more general model,
while line $VD$ is the well-studied monomer-dimer model~\cite{Kastelyn,Temperly_Fisher,Fisher,Fisher_Stephenson,Heilmann_Lieb,Nagle_Okoi_Bhattacharjee,Huse_etal,Kenyon_Okounkov_Sheffield}. For the vacancy-free mixure along line $DS$ (Fig.~\ref{Fig1} b), we show that the power-law columnar order present in the dimer limit $D$ is enhanced by adding hard-squares. This eventually leads to a Kosterlitz-Thouless (KT) phase transition from this power-law ordered phase to a hard-squares-rich phase with long-range columnar order (Fig.~\ref{Fig1} c). Noting that the power-law ordered phase and the KT point
are both characterized by an emergent U($1$) symmetry, we show that correlations of the two-sublattice
order parameter of hard-squares decay in this regime with the same power law exponent as those of the nematic order parameter. With vacancies allowed, we establish that the phase boundary (Fig.~\ref{Fig1} b)  between
this columnar ordered phase and the low-density fluid is 
in the Ashkin-Teller (AT) universality class with a fixed anomalous
exponent $\eta=1/4$ for the columnar order parameter, and a continuously varying correlation length exponent $\nu$. We
also demonstrate that the anomalous exponent $\eta_2$ for {\em nematic} order 
obeys an {\em Ashkin-Teller
relation} $\eta_2 = 1-1/(2\nu)$ along the phase boundary, {\em including at the hard-square transition}, thus settling the original question of critical properties at the hard-square transition.  These results are made
possible by our identification of a detailed correspondence between the microscopic hard-square and dimer variables measured
in our Monte-Carlo simulations and the $XY$ (Ising) order-parameter fields of a long-wavelength description of KT (AT) criticality.

{\em Model:} Our analysis begins by defining a lattice-gas~(Fig.~\ref{Fig1} a) of hard-squares that occupy the four  elementary plaquettes of a square lattice, dimers that occupy two plaquettes, and vacant single plaquettes (vacancies/monomers). 
We consider a $L \times L$ square lattice with periodic boundary conditions
and associate activities $z_s, z_d$ and $z_v$ with each square, dimer and vacancy respectively. The grand partition function of the system is then given by
\begin{equation}
Z_{dsv}= \sum_{{\mathcal C}_{dsv}}  z_s^{N_{s}}z_d^{N_{d}}z_v^{N_v} \; .
\end{equation}
Here, the sum is over all  allowed configurations ${\mathcal C}_{dsv}$ that respect the
hard-core constraints  (Fig.~\ref{Fig1} a), and  $N_s$, $N_{d}$ and $N_{v}$, the total numbers of squares, dimers and vacancies, obey the constraint  $4N_{s} + 2N_{d} + N_v = L^2$,
allowing us to parametrize results in terms of two independent parameters: $v = z_vz_s^{-1/4}$, and $w = z_d/\sqrt{z_s}$.

{\em Line $DS$: } At $v=0$, $Z_{dsv}$ reduces to $Z_{ds}$, the partition function of a vacancy-free
mixture of squares and dimers. In the $z_s \rightarrow 0$ limit, $Z_{ds}$ further
reduces to $Z_{{\mathrm{dimers}}}$, the partition function of the fully-packed dimer model.
$Z_{{\mathrm{dimers}}}$ is characterized by a power-law tendency to columnar order
manifested in the connected correlation function of
horizontal (vertical) dimers, which decays as $(-1)^l/l^2$ for large separation $l$ 
along  the $x$ ($y$) axis~\cite{Fisher_Stephenson}. For small but non-zero $w^{-1}$, $Z_{ds}$ involves configurations with a small density
of squares. Regarding each square as a length-four loop and each dimer as a length-two loop on the dual lattice allows us to use the recursive procedure of Ref.~\cite{Damle_Dhar_Ramola} to map $Z_{ds}$ to an interacting dimer-model with $k$-dimer interactions ($k=2,3 \dots$).
The leading interaction is a two-body attraction $V_2$ of strength $\log[1+1/(2w^2)]$ between two adjacent dimers whose long sides touch fully.
As seen in earlier work~\cite{Alet_etal_PRL,Papanikolaou_Luijten_Fradkin,Alet_etal_PRB}, this interaction enhances the power-law columnar order present
in the dimer-limit, with power-law exponent $\eta(w)$ decreasing from $\eta(w=\infty) = 2$
as $V_2$ increases in strength. 
Furthermore, the net effect of the $k>2$ interaction terms
also favours columnar ordering.  Therefore,
for $w$ less than a critical value $w_c^{(0)}$, we expect a phase with long-range columnar order. In this columnar state, the symmetry of $\pi/2$ rotations is broken and the unit-cell is doubled in the direction perpendicular to the
stripes that form (Fig.~\ref{Fig1} c). 

This four-fold symmetry-breaking is conveniently characterized in
terms of a complex order parameter $\psi(\vec{r})$ defined on plaquettes
$\vec{r}$ in terms of microscopic variables as follows: $\psi(\vec{r})$ vanishes at $\vec{r}$ if  plaquette $\vec{r}$ is vacant. Otherwise, it takes on
the values depicted in Fig.~\ref{Fig1} a). These values are specified based on the coordinate $\vec{R} \equiv (m,n)$ of the bottom, left corner of the tile covering $\vec{r}$ 
as follows:
\begin{equation}
\psi_1 = (-1)^m , \; \psi_2 = -i (-1)^n , \;  \psi_3 =[ (-1)^m - i (-1)^n]/\sqrt{2} \; .
\label{psidefinition}
\end{equation} 
With this definition, $\langle \psi \rangle$ takes on values  $ \pm a, \pm i a$  in the four symmetry-related columnar-ordered states (the magnitude $a>0$ depends on the composition of the mixture), while $\langle \psi^{*}(\vec{r})\psi(0)\rangle$ falls off as $1/r^{\eta(w)}$ for large $r$ in the power-law columnar-ordered phase.

To understand the nature of the transition at $w_c^{(0)}$ along $DS$ (Fig.~\ref{Fig1} b),  we
use the fact that
$Z_{ds}$ admits a height representation, {\em i.e.} the microscopic configurations are
uniquely specified in terms of a single-valued scalar height
$H(\vec{R})$ defined on lattice sites $\vec{R}$ as follows: Set $\eta_{mn} \equiv (-1)^{m+n}$ and the height at the
origin $H(\vec{O})=0$. To construct the height field $H(\vec{R})$, traverse any sequence of
links of the square lattice to go from $\vec{O}$ to $\vec{R} \equiv (m,n)$. When traversing a vertical
link from $(m,n)$ to $(m,n+1)$ (horizontal link from $(m+1,n)$ to $(m,n)$),  $H$ increases by $3\eta_{mn}/4$  if this link is fully
covered by a dimer, by $\eta_{mn}/4$ if fully covered by a square, and  by   $-\eta_{mn}/4$ otherwise.   When there are no squares, this reduces to the well-known height representation for the fully-packed dimer model~\cite{Alet_etal_PRB,Youngblood_Axe_McCoy,Youngblood_Axe,Blote_Hillhorst,Nienhuis_Hillhorst_Blote,Kondev_Henley_PRB, Zeng_Henley,Raghavan_Henley_Arouh, Fradkin_Huse_Moessner_Oganesyan_Sondhi}.

In the $w> w_c^{(0)}$ power-law ordered phase, long-wavelength fluctuations of the height-field are well-described by the effective action~\cite{Alet_etal_PRB,Youngblood_Axe_McCoy,Youngblood_Axe, Blote_Hillhorst,Nienhuis_Hillhorst_Blote,Zeng_Henley, Fradkin_Huse_Moessner_Oganesyan_Sondhi}:
\begin{eqnarray}
S_{\rm eff}= \int_{\Lambda} d^2x \left [\pi g(\nabla h)^2 + \hspace{-0.2cm} \sum_{n=4,8,12 \dots} \hspace{-0.3cm} u_n \cos(2\pi nh) \right].
\end{eqnarray}
Here $h$ is a coarse-grained version of the microscopic height field $H(\vec{R})$, the 
values of the stiffness $g$ and  $n$-fold anisotropy terms $u_n$ at the coarse-graining scale $\Lambda$ are phenomenological parameters, and the form of the cosine terms in the action are fixed~\cite{Alet_etal_PRB,Zeng_Henley,Fradkin_Huse_Moessner_Oganesyan_Sondhi} by the transformation properties of
$h$ under lattice-symmetries of the original partition function.

The utility of $S_{\rm eff}$ lies in two observations:
First, since $e^{2\pi i h(\vec{r}) }$ transforms~\cite{Alet_etal_PRB,Zeng_Henley,Fradkin_Huse_Moessner_Oganesyan_Sondhi} under lattice-symmetries in the same way
as $\psi(\vec{r})$, we expect long-distance properties of correlators of
$\psi(\vec{r})$ in $Z_{dsv}$ to correspond to those of $e^{2 \pi i h(\vec{r})}$ in
the coarse-grained theory $S_{\rm eff}$. Second,
$S_{\rm eff}$ with all $u_n$ set to zero represents a line of critical fixed-points parameterized by a variable stiffness $g$. All allowed cosine terms $u_n$ are
irrelevant perturbations of this fixed line for $g < 4$~\cite{Jose_Kadanoff_Kirkpatrick_Nelson}.  Along
this fixed line~\cite{Jose_Kadanoff_Kirkpatrick_Nelson}, 
$\langle e^{2\pi i (h(\vec{r})-h(0))} \rangle$ falls off as $1/r^{1/g}$.
This implies power-law columnar order with exponent $\eta = g^{-1}$, since correlations
of $\psi(\vec{r})$ and $e^{2\pi i h(\vec{r})}$ have the same long-distance behaviour.
Therefore, we may identify the $w \rightarrow \infty$ limit of $Z_{ds}$ with
the point~\cite{Alet_etal_PRB} $g = 1/2$ on this fixed line, consistent with $\eta(\infty) = 2$. Since we have already
argued that $\eta(w)$ reduces as $w^{-1}$ is increased from $0$, we expect that
the corresponding value of $g$ increases on this fixed line until it hits $g=4$, corresponding
to $\eta = 1/4$. At this point, $u_4$ becomes marginally
relevant, driving a Kosterlitz-Thouless (KT) transition to a four-fold symmetry-breaking state with long range order for $e^{2 \pi i h(\vec{r})}$, {\em i.e.} a columnar ordered state with nonzero $\langle \psi \rangle$.  

This irrelevance of all cosine terms in
the power-law ordered phase implies
that the phase of $\Psi_L \equiv \sum_r \psi(\vec{r})$ for large $L$ will be {\it uniformly distributed } in $(0,2\pi)$ throughout the power-law ordered phase and at the KT point, reflecting the presence of an emergent $U(1)$ symmetry. From their microscopic
expressions, we note that ${\rm Re}(\psi^2(\vec{r}))$
measures nematic order in terms of orientations of dimers,
while ${\rm Im}(\psi^2(\vec{r}))$ 
is the two-sublattice
order-parameter of hard-squares.
This $U(1)$ symmetry implies that $\eta_{s}$,
the anomalous exponent governing the power-law correlations of ${\rm Im} (\psi^2(\vec{r}))$,
equals
$\eta_2$, the corresponding exponent for ${\rm Re} (\psi^2(\vec{r}))$.
The Gaussian nature of height-fluctuations
further ensures that both $\eta_2$ and $\eta_{s}$ equal $4\eta$ throughout this power-law phase and at the KT point.

\begin{figure}
{\includegraphics[width= 0.85\columnwidth]{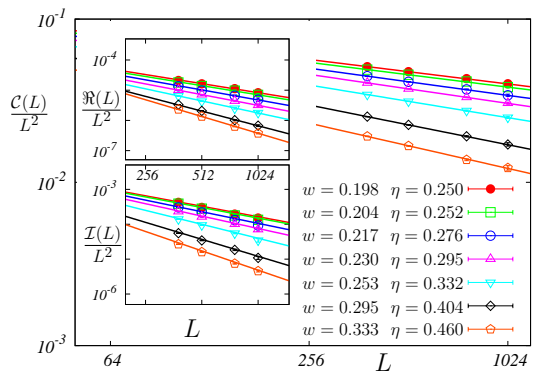}}
\caption{(Color online) 
$\mathcal{C}(L)/L^2 \sim L^{-\eta(w)}$ with variable exponent $\eta(w)$ in the power-law ordered phase at full-packing. 
Insets: $\Re(L)/L^2$
and  $\mathcal{I}(L)/L^2$
both scale as $L^{-4\eta(w)}$ with the same $\eta(w)$.}
\label{Fig2}
\end{figure}

{\em The AT phase-boundary: } 
The KT transition at $(w=w_c^{(0)},v=0)$ represents the begining of an
Ashkin-Teller (AT) critical line in the phase-diagram of $Z_{dsv}$ (Fig.~\ref{Fig1} b), at whose
other end $(w=0,v=v_c^{*})$ lies the density-driven transition of the hard-square lattice
gas. To establish this, we first note that it is enough
to keep a non-zero $u_4$ and set all other $u_n$ in $S_{\rm eff}$ to zero in the vicinity of this KT transition at $g=4$~\cite{Jose_Kadanoff_Kirkpatrick_Nelson}. Thus, the $v=0$ KT transition can be thought of as a transition
to long-range order in a vortex-free $XY$ model with four-fold anisotropy.  Next, we note that
an isolated vacancy on plaquette $\vec{r} = (m+1/2,n+1/2)$ causes the phase of  the $XY$ order parameter $\psi(\vec{r}) $ to wind by $2\pi \times (-1)^{m+n}$ along
a circuit that encloses the vacant plaquette once. On the vacant plaquette
itself, $\psi=0$, as befits the core of a vortex in an $XY$ order parameter.
 Thus, a non-zero density of vacancies in $Z_{dsv}$ corresponds to 
perturbing this vortex-free, four-fold anisotropic $XY$ model with a non-zero density of vortices and anti-vortices. 
As is well-known from  the work of Kadanoff and others on such $XY$ models with four-fold anisotropy~\cite{Jose_Kadanoff_Kirkpatrick_Nelson,Kadanoff_PRL,Kadanoff_JPhysA,Kadanoff_Annals,Kadanoff_Brown,Cardy,Rujan_Williams_Frisch_Forgacs,Alet_etal_PRL,Alet_etal_PRB,Papanikolaou_Luijten_Fradkin}, vorticity and four-fold anisotropy ``balance'' each
other along a  line of fixed points that starts at this vortex-free KT point. This fixed line
describes the continuously-varying critical properties of the Ashkin-Teller (AT) universality class~\cite{Ashkin_Teller,Baxter_PRL,Kadanoff_Wegner,Jose_Kadanoff_Kirkpatrick_Nelson,Kadanoff_PRL,Kadanoff_JPhysA,Kadanoff_Annals,Kadanoff_Brown,Elitzur_Pearson_Shigemitsu,Cardy,Rujan_Williams_Frisch_Forgacs,Kohmoto_denNijs_Kadanoff,Boyanovsky,Delfino,Alet_etal_PRL,Alet_etal_PRB,Papanikolaou_Luijten_Fradkin,Taroni_Bramwell_Holdsworth}, {\em i.e.} the critical behaviour of two Ising models coupled via their energy-densities. 
\begin{figure}
{\includegraphics[width=0.85\columnwidth]{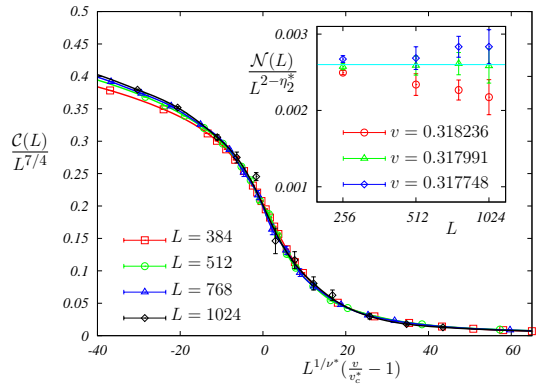}}
\caption{(Color online) Scaling collapse of $\mathcal{C}(L)/L^{7/4}$ for various $L$ at the hard square transition yields
the estimate $\nu^{*} = 0.92(3)$  and $v_c^{*}= 0.31799(30)$.
Inset: $\mathcal{N}(L)/L^{2-\eta_2^{*}}$ is a constant
for $v=v_c^{*}$ with $\eta_2^{*} \approx 0.46(3)$.}
\label{Fig3}
\end{figure}
For $Z_{dsv}$, this implies that
the $(w=w_c^{(0)},v=0)$ KT transition represents the start of an AT critical
line that separates a square-rich columnar-ordered phase from
a low-density fluid phase (Fig.~\ref{Fig1} b). The density-driven transition at $(w=0,v=v_c^*)$ in the hard-square lattice gas thus represents the other end of this AT line. The two real scalar fields $\sigma$
and $\tau$ of this alternate Ashkin-Teller description are defined in terms of the $XY$  order
parameter $\psi$ (defined in Fig.~\ref{Fig1} a) by the equation
\begin{equation}
\psi(\vec{r}) \equiv \frac{\sigma(\vec{r})+ \tau(\vec{r})}{2} + i \frac{\sigma(\vec{r})- \tau(\vec{r})}{2}.
\end{equation}
From their expressions in terms of microscopic variables, it
is clear that lattice symmetries only guarantee
\begin{equation}
\langle \sigma(\vec{r}_{1}) \tau(\vec{r}_{2}) \rangle = 0 \; , \; \langle \sigma(\vec{r}) \sigma(0)\rangle = \langle \tau(\vec{r}) \tau(0) \rangle .
\end{equation} 
In particular, $\langle \sigma^2(\vec{r}) \tau^2(0)\rangle$ is not constrained to vanish even in the pure hard-square limit, and there is no symmetry reason to expect that the Ising fields
$\sigma$ and $\tau$ are asymptotically decoupled. 

\begin{figure}
{\includegraphics[width=0.85\columnwidth]{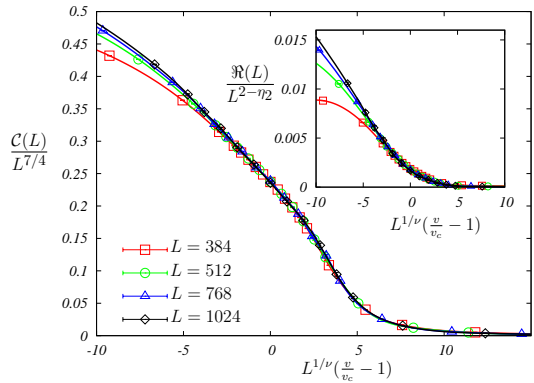}}
\caption{(Color online) Scaling collapse of $\mathcal{C}(L)/L^{7/4}$ for various $L$ along a cut
that crosses the AT boundary at an intermediate point $w_c=0.1600(1)$, $v_c=0.0623(1)$ yields
the estimate $\nu = 1.70(5)$.
Inset: Scaling collapse of $\Re(L)/L^{2-\eta_2}$ yields the estimate $\eta_2 \approx 0.70(5)$.}
\label{Fig4}
\end{figure}

{\em Numerics: } 
 These ideas, in conjunction with our knowledge~\cite{Jose_Kadanoff_Kirkpatrick_Nelson,Kadanoff_PRL,Kadanoff_JPhysA,Kadanoff_Annals,Kadanoff_Brown,Cardy,Rujan_Williams_Frisch_Forgacs,Alet_etal_PRL,Alet_etal_PRB,Papanikolaou_Luijten_Fradkin} 
of the long-wavelength physics of the Ashkin-Teller universality class, lead to three key
predictions that can be tested via numerical simulations:
All along the AT phase boundary, $\langle \psi^{*}(\vec{r}) \psi(0)\rangle$ is predicted to falls off as $1/r^{1/4}$, while
$\langle {\rm Re}(\psi^2(\vec{r})){\rm Re}(\psi^2(0))\rangle$ is expected to decay as $1/r^{\eta_2(v)} $,
where $\eta_2(v)$ varies continuously, starting from the $v=0$ value $\eta_2(v=0) = 1$.
Thus $\eta_2$ is a natural coordinate in terms of which
one can specify the position along the AT phase-boundary. Moreover, the correlation-length exponent $\nu$ is related to
$\eta_2$ via an {\em Ashkin-Teller relation}: 
\begin{equation}
\eta_2=1-1/(2\nu) .
\end{equation} 

In the power-law ordered phase at full-packing, our earlier results imply, via
finite-size scaling, that $\mathcal{C}(L)= \langle |\Psi_L|^2\rangle/L^2 $  scales
as $L^{2-\eta(w)}$, while $\Re(L) = \langle [\sum_{\vec{r}} {\rm Re} (\psi^2(\vec{r}))]^2\rangle/L^2 $ and $ \mathcal{I}(L)= \langle [\sum_{\vec{r}} {\rm Im} (\psi^2(\vec{r}))]^2\rangle/L^2$
scale as $L^{2-4\eta(w)}$. In the vicinity of the AT phase boundary, finite-size scaling implies that
$\mathcal{C}(L)$  and $\Re(L)$  are expected to satisfy the scaling forms $L^{7/4}f_{\mathcal{C}}(\delta L^{1/\nu})$  and $L^{2-\eta_2(v)}f_{\Re}(\delta L^{1/\nu})$ respectively, where $\delta$ denotes the deviation from criticality and the $f$ are finite-size scaling functions.
Close to the density-driven hard-square transition, it is more convenient to measure $\eta_2$ 
using an alternate nematic
order parameter $T(\vec{r})$ which keeps track of the {\em orientations of vacancy-pairs and dimers adjacent to hard-squares}: $T(\vec{r})=0$ when $\vec{r}$ is not covered by a hard-square. Otherwise
$T(\vec{r})  \equiv T_H(\vec{r}) - T_V(\vec{r})$, where $T_H(\vec{r})$ ($T_V(\vec{r})$) equals one-quarter the total number of horizontal (vertical) vacancy-pairs or dimers immediately adjoining the hard square that covers $\vec{r}$. $T(\vec{r})$  transforms in the same way as ${\rm Re}(\psi^2(\vec{r}))$, and $\langle T(\vec{r}) T(0)\rangle$ is predicted to
also decay as $1/r^{\eta_2}$ at criticality. By finite-size scaling, this implies
that $\mathcal{N}(L) \equiv \langle (\sum_{\vec{r}}T(\vec{r}))^2\rangle/L^2$ is expected
to have the scaling form $L^{2-\eta_2}f_{\mathcal{N}}(\delta L^{1/\nu})$ in the vicinity of the hard-square transition.

To test these predictions, we have performed Monte-Carlo simulations
of $Z_{dsv}$ on $L\times L$ periodic lattices (with $L$ upto $1024$) using a
variation~\cite{Supplemental} of an algorithm~\cite{Kundu_Rajesh_Dhar_Stilck} which generates, in a single move, an equilibrium configuration of an entire
row (or column), given the configuration of the rest of the system. Our method does not suffer from jamming even at full-packing, and can be generalized to a large class of similar problems. More details are provided in the Supplemental Material~\cite{Supplemental}.
For $w > w_c^{(0)} \approx 0.198(2)$ along $DS$, we find clear evidence of a $v=0$
power-law ordered phase, in which $\mathcal{C}(L)/L^2$ decays as $1/L^{\eta(w) }$, while $\Re(L)/L^2$ and $\mathcal{I}(L)/L^2$  both decay as $1/L^{4\eta(w)}$,
with $\eta(w_c^{(0)})= 1/4$ (Fig.~\ref{Fig2}). For the hard-square lattice gas, we estimate that
the transition point is located at $v_c^* = 0.3180(3)$. Our data for ${\mathcal C}(L)$
is well-fit by $\eta=1/4$, and $\nu^{*} \approx 0.92(3)$, consistent with some of the earlier studies~\cite{Zhitomirsky_Tsunetsugu,Feng_Blote_Nienhuis}, while $\mathcal{N}(L)$ diverges as $L^{2-\eta_2^{*}}$ at criticality, with $\eta_2^{*} \approx 0.46(3)$ (Fig~\ref{Fig3}), consistent with the Ashkin-Teller relation, providing conclusive evidence of the AT nature
of the hard-square transition, and emphasizing that the hard-square transition
lies beyond the decoupled Ising point (Fig.~\ref{Fig1} b) on the AT phase boundary.
Additionally, at an intermediate point (Fig.~\ref{Fig1} b) on the phase-boundary, our data for ${\mathcal C}(L)$ is fit
well by $\eta=1/4$ and $\nu \approx 1.70(5)$, while $\Re(L)$ grows as $L^{2-\eta_2}$ at criticality, with  $\eta_2  \approx 0.70(5)$ (Fig.~\ref{Fig4}), consistent with the Ashkin-Teller relation. This provides the first test of this relation in a microscopic lattice model with
continuously varying exponents.

{\em Outlook:} Given that columnar ordering is ubiquitous in a wide variety
of strongly-correlated systems~\cite{Jin_Sandvik,Wenzel_Coletta_Korshunov_Mila,Sen_Damle_Senthil,Edlund_Jacobi,Jin_Sen_Sandvik,Ralko_Poilblanc_Moessner},
the ideas discussed here are of immediate relevance in a variety of other contexts. For instance,  the emergent U($1$) symmetry at full-packing is closely related to the U($1$) symmetry
that is expected to emerge in the zero temperature limit~\cite{Senthil_etal_Science,Senthil_etal_PRB} of the thermal AT transition~\cite{Jin_Sandvik}  to columnar valence-bond solid (VBS)
order in a class of frustrated square-lattice antiferromagnets that have been the focus of many recent studies~\cite{Sandvik_PRB2012,Sandvik_PRL2010,Sandvik_PRL2007,Banerjee_etal_2010,Kaul_2011,Banerjee_etal_2011,Lou_etal_PRB09,Kaul_Sandvik_PRL2012,Melko_Kaul_PRL2008,Jiang_etal_JStatmech2008,Chen}. 
The ideas developed here predict that this emergent U($1$) symmetry 
constrains the behaviour of certain subdominant orders at this ``deconfined' quantum
critical point~\cite{Senthil_etal_Science,Senthil_etal_PRB}.  More precisely, with $\psi(\vec{r})$ now representing
the complex VBS order parameter,  we predict that correlations of ${\rm Re}(\psi^2(\vec{r}))$, the valence-bond nematic order parameter, decay with power-law exponent $\eta_{\rm VBN}$ that equals  the power-law decay exponent for correlations
of ${\rm Im}(\psi^2(\vec{r}))$, the wavevector $(\pi,\pi)$ component
of the next-nearest-neighbour bond-energy, at this quantum critical point. Additionally, we predict that $\eta_{\rm VBN}$ and $\nu$, the correlation length exponent for VBS order
parameter correlations, are related all along the AT phase boundary via the Ashkin-Teller relation discussed here.

\textit{Acknowledgements}
We gratefully acknowledge useful comments by M. Barma on an earlier draft
of our manuscript. This research was
supported by the Indo-French Centre for the Promotion of Advanced Research (IFCPAR/CEFIPRA) under Project 4603-3 (DD), and by the Indian DST via grant DST-SR/S2/JCB-24/2005 (DD). We gratefully acknowledge use of computational resources funded by DST grant
DST-SR/S2/RJN-25/2006 (KD), in addition to departmental computational
resources of the Dept. of Theoretical Physics of the TIFR. Some
of our results on the hard-square lattice-gas were summarized earlier in the doctoral thesis~\cite{thesis} of K.~Ramola at the TIFR.

\clearpage

\begin{widetext}

\begin{appendix}
\section*{\large Supplemental Material for ``Columnar order and Ashkin-Teller criticality in
mixtures of hard-squares and dimers"}

In this document we present details of our Monte Carlo algorithm
and additional results from our simulations which support
the key findings highlighted in the main text.

\section{Transfer-Matrix Based Algorithm}

\subsection{Update Scheme}

To simulate the system of dimers and squares on the square lattice, we use the following transfer-matrix 
based Monte Carlo algorithm
which is a variant of the technique developed in
Refs.~\cite{Kundu_Rajesh_Dhar_Stilck,Kundu_etal,thesis,Kundu_Rajesh,Nath_Rajesh}. 
Our variant is designed to ensure
that we can work directly in the full-packing limit if needed.
In our scheme, we update all objects fully contained in a $2 \times N$ track (two adjacent rows/columns of plaquettes) 
at once, with the correct weights in the partition function. 
The steps involved in each update are as follows:

\begin{itemize}
\item We empty out all objects that are fully contained within a randomly chosen $2\times N$ track (horizontal or vertical).

\item The remaining objects either lie outside the chosen track (this includes
objects which share an edge with the long boundary of the track) or protrude partially
into the track.  The latter class of objects, which protrude partially into
the track, provide excluded-volume constraints that need to be respected
when the track is refilled.

\item To refill the track with objects lying entirely within the track, we
compute the partition function of the track subject to the constraints imposed
by objects that protrude into the track. This is done using a standard transfer matrix technique.                                                                 

\item Using this partition function, we generate a configuration with the correct
Botzmann weight consistent with the constraints, and re-populate the track.
We summarize these steps in Fig. \ref{update_fig} below.
\end{itemize}
\vspace{-0.2cm}
\begin{figure}[h!]
{\bf 1.}
\includegraphics[width=.44\columnwidth,angle=0]{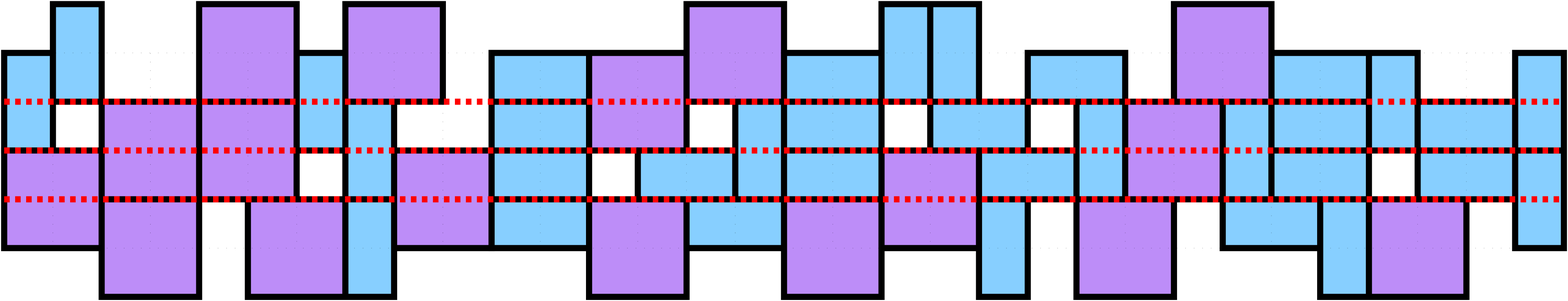}\\
\vspace{0.35cm}

{\bf 2.}
\includegraphics[width=.44\columnwidth,angle=0]{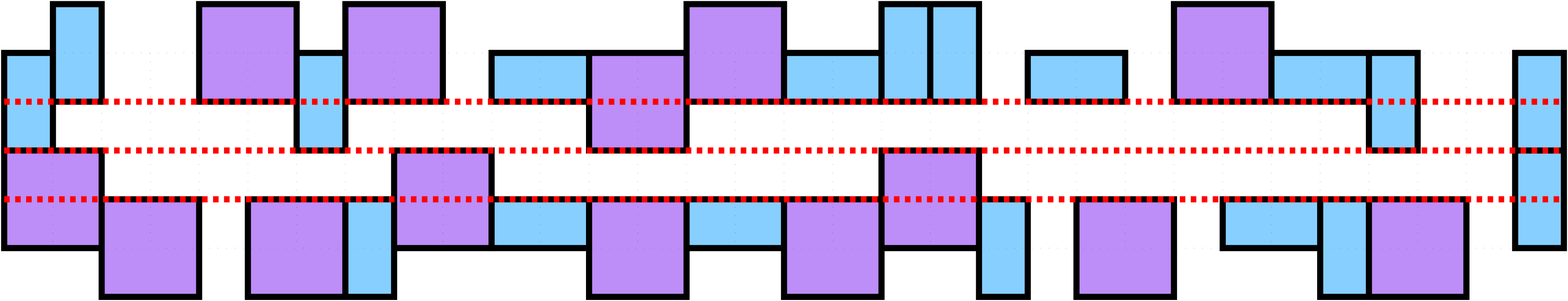}\\
\vspace{0.35cm}

{\bf 3.}
\includegraphics[width=.44\columnwidth,angle=0]{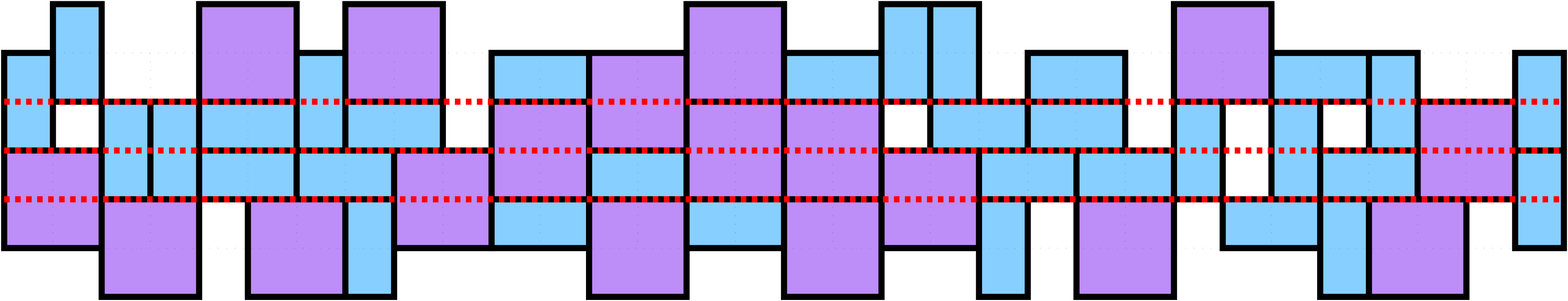}
\caption{Steps in the transfer-matrix based algorithm. {\bf 1.} The initial configuration of the randomly
chosen $2 \times N$ track (red dashed lines), showing objects lying completely within it, objects lying immediately outside it
(but sharing an edge with the track), and objects protruding into the track.
{\bf 2.} All objects lying fully within the track are evaporated. We leave
unchanged all objects that lie completely outside the track (not shown)
or objects lying immediately outside
the track (sharing an edge with the track without protruding into the track),
as well as objects that protrude into the track.
{\bf 3.} The track is re-populated with a new configuration of objects lying entirely
within the track, subject to the excluded volume constraints imposed by objects
protruding into the track.}
\label{update_fig}
\end{figure} 

To evaluate the weights of the allowed configurations for the purpose
of refilling a track, we need to calculate 
the restricted partition function of this track subject to constraints imposed by
objects protruding into the track. 
We do this by using a standard transfer matrix technique. 
Below we provide details of this update for a horizontal track.

\subsection{Details of the Transfer Matrices}

We break up the track into a sequence of two-plaquette
``rungs'', defined as two vertically adjacent plaquettes.
After the track is emptied of all objects lying completely within it,
these rungs still have areas covered by objects protruding into the track from above and below 
(as shown in {\bf 2.} of Fig. \ref{update_fig}).
These protrusions preclude the occupation of some objects on the rung, and thereby provide constraints on which
objects can be re-populated.
The four possible types of protrusions (represented by shaded areas) on a given rung are shown in Fig. \ref{rung_blocks_fig}.
Based on this underlying ``morphology'', we assign an index $\sigma$ to each rung, with $\sigma = 1,2,3,4$ chosen with  
the convention of Fig. \ref{rung_blocks_fig}.


\begin{figure}[h!]
{\bf 1.}
\includegraphics[width=.06\columnwidth,angle=0]{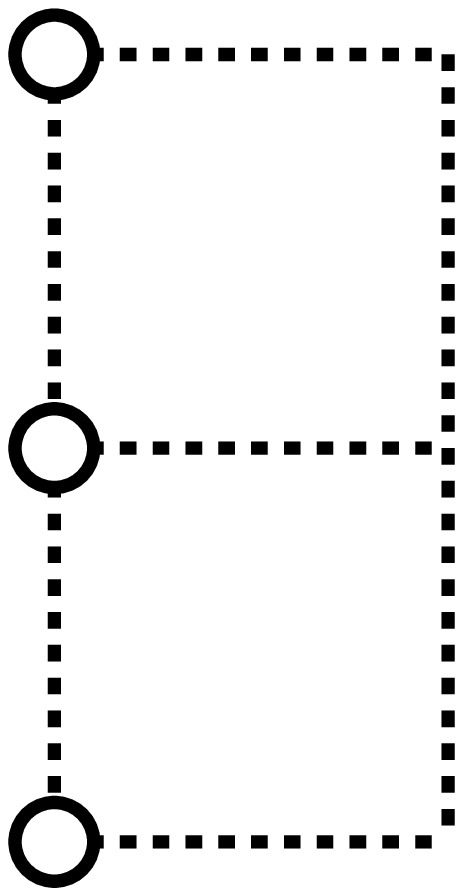}
\hspace{0.5cm}
{\bf 2.}
\includegraphics[width=.06\columnwidth,angle=0]{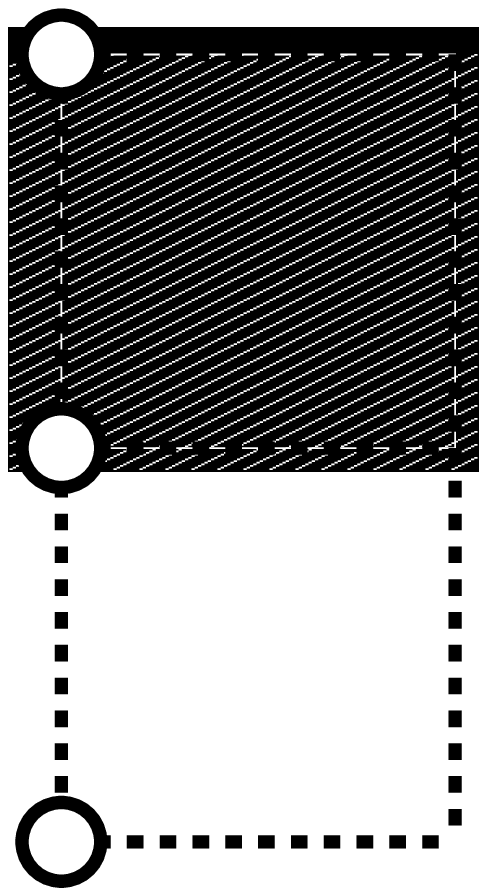}
\hspace{0.5cm}
{\bf 3.}
\includegraphics[width=.06\columnwidth,angle=0]{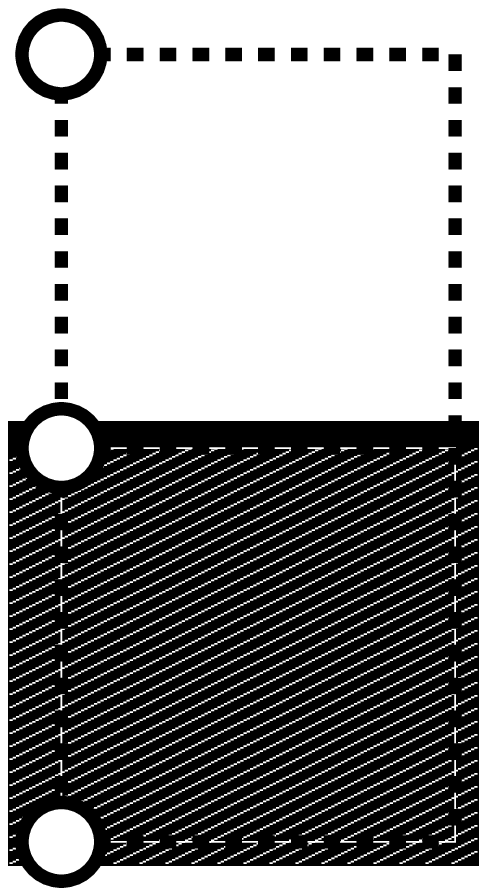}
\hspace{0.5cm}
{\bf 4.}
\includegraphics[width=.06\columnwidth,angle=0]{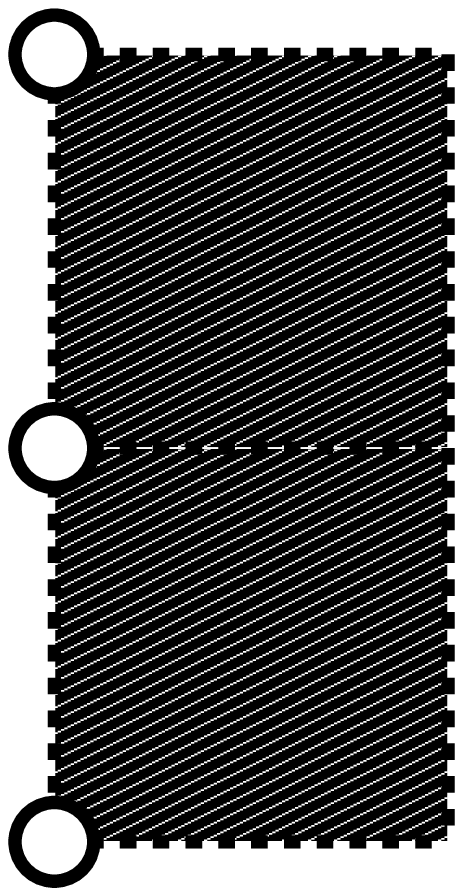}
\caption{The four possible underlying morphologies $\sigma = 1,2,3,4$ of a two-plaquette rung,
arising from objects protruding into the track from above and below (represented by the shaded areas).
$\sigma = 4$ corresponds to a complete blockade.}
\label{rung_blocks_fig}
\end{figure}

Next, in order to fill the rung with objects,
we focus on the ``state'' $C$ of a rung, the ways in which objects can be 
placed on this rung. 
When the underlying morphology is ignored, 
there are {\bf six} possible ways of filling a two-plaquette rung, 
as shown in Fig. \ref{states_fig}.

\begin{figure}[h!]
{\bf 1.}
\includegraphics[height=.1\columnwidth,angle=0]{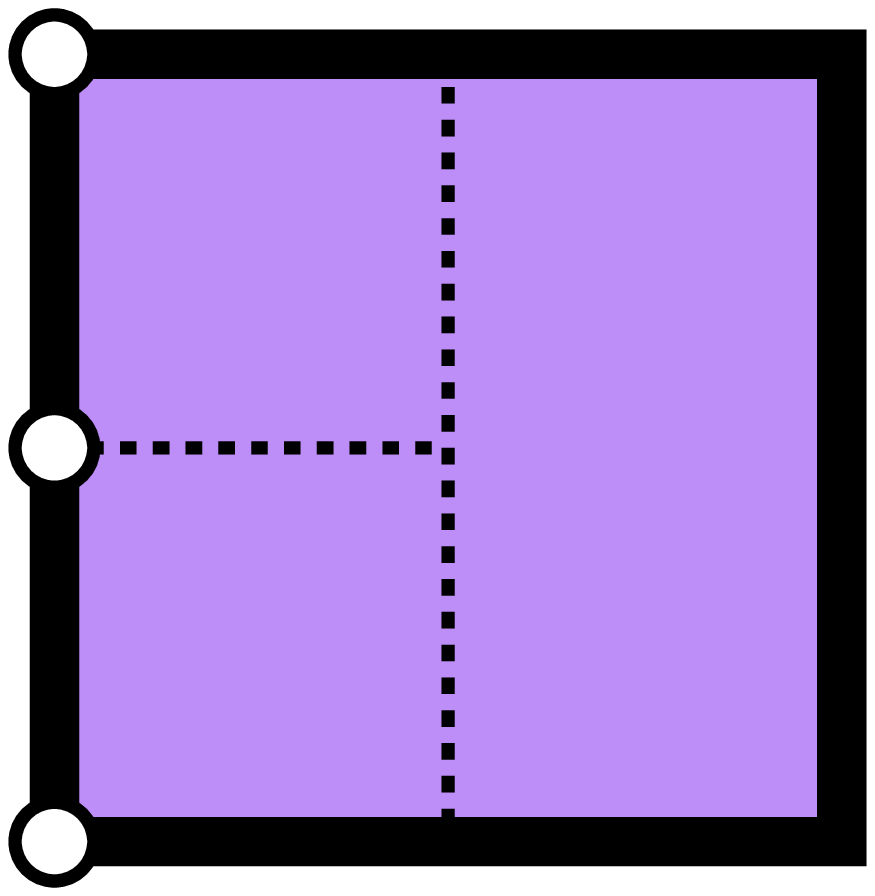}
\hspace{0.5cm}
{\bf 2.}
\includegraphics[height=.1\columnwidth,angle=0]{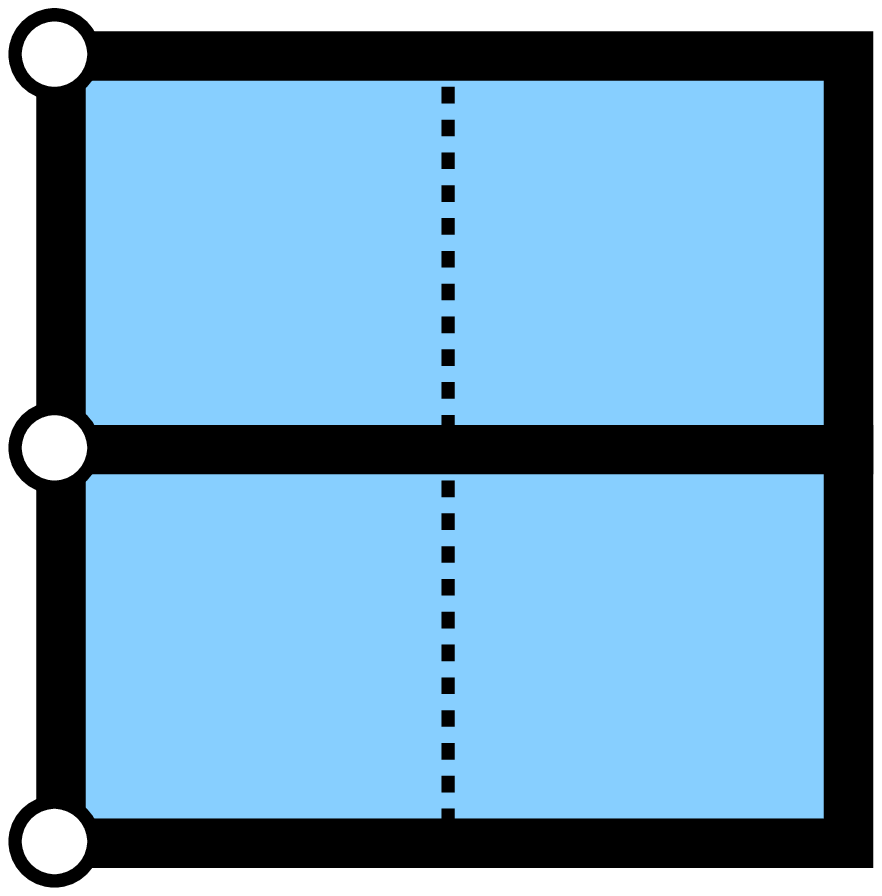}
\hspace{0.5cm}
{\bf 3.}
\includegraphics[height=.1\columnwidth,angle=0]{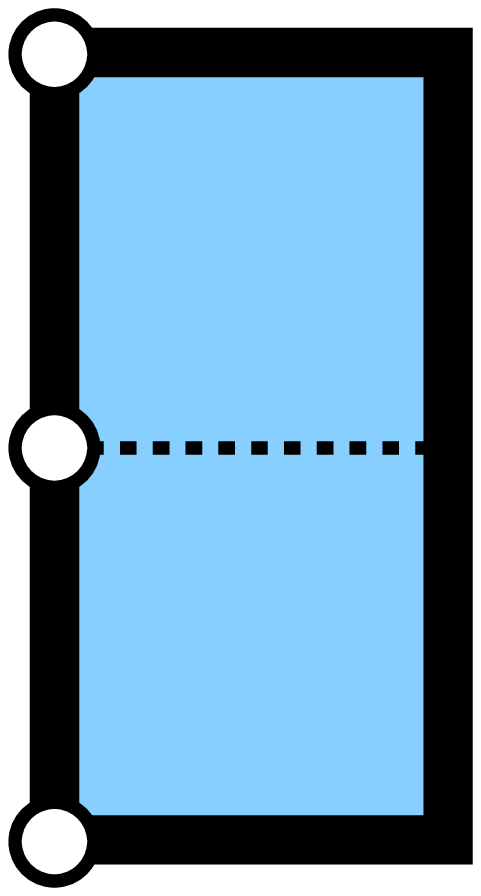}
\hspace{0.5cm}
{\bf 4.}
\includegraphics[height=.1\columnwidth,angle=0]{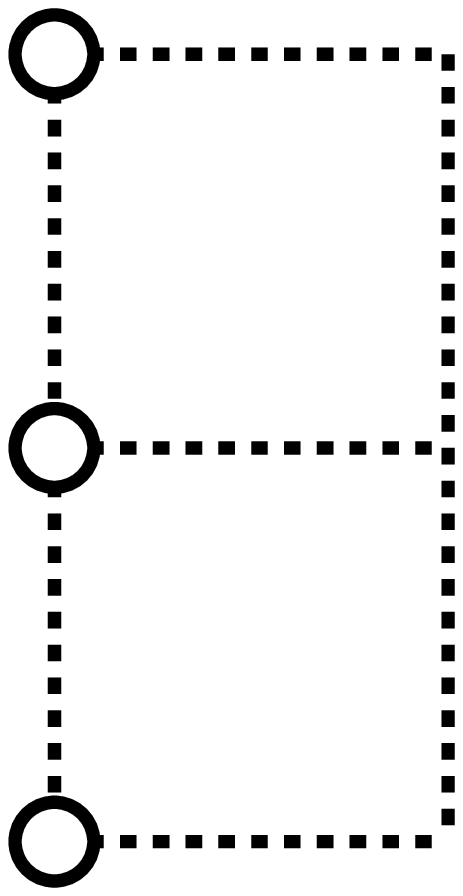}
\hspace{0.5cm}
{\bf 5.}
\includegraphics[height=.1\columnwidth,angle=0]{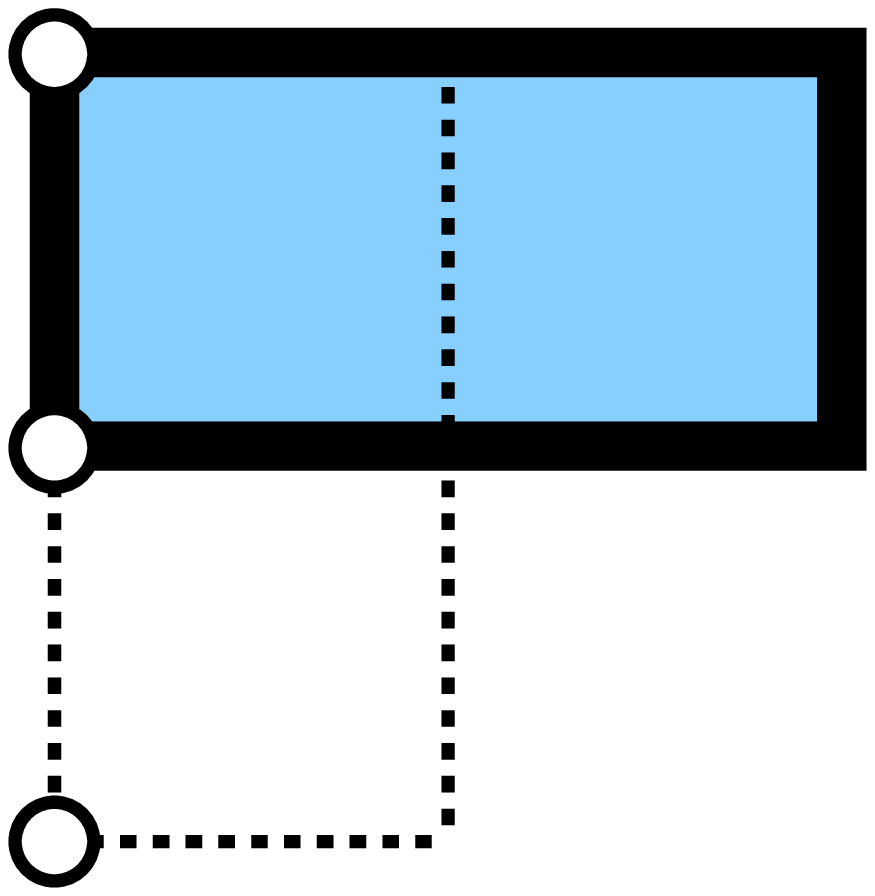}
\hspace{0.5cm}
{\bf 6.}
\includegraphics[height=.1\columnwidth,angle=0]{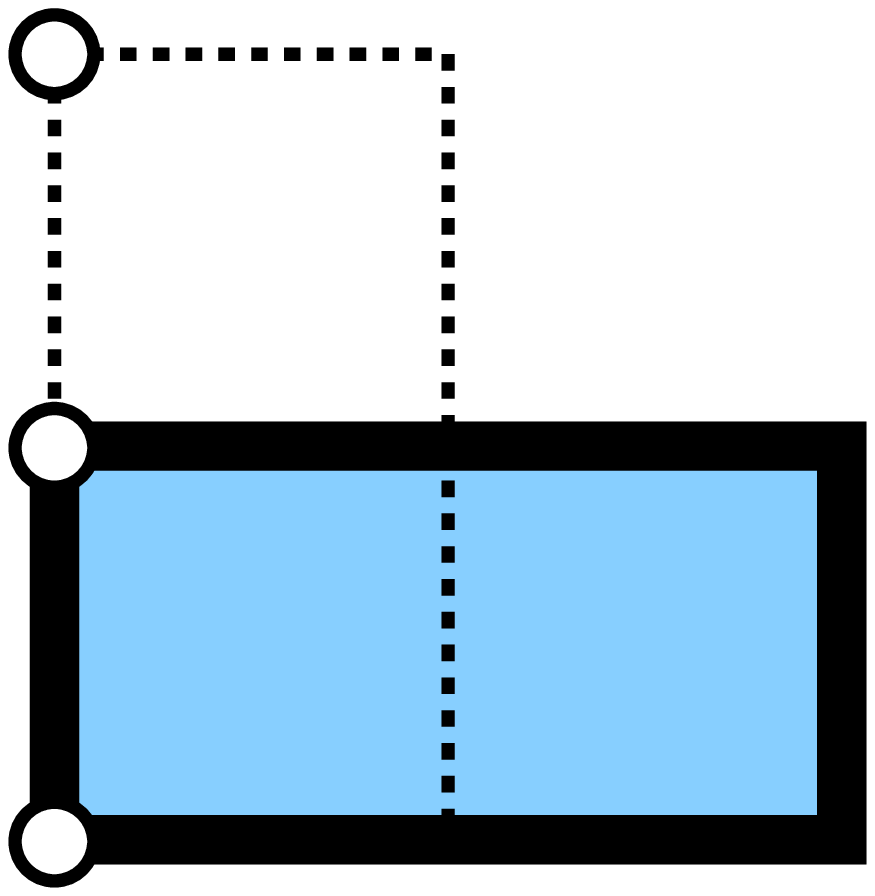}
\caption{The six possible states of a two-plaquette rung.}
\label{states_fig}
\end{figure}

To unambiguously assign objects to each rung, we use 
the convention that objects are on the rung, if their 
left edge coincides with the left edge of the
rung (represented by open circles in Fig.~\ref{states_fig}).
Our convention is also designed to ensure that the allowed
states are influenced {\it only} by the morphology of the given rung
and the one immediately to the right.
When the underlying morphology is considered, not all states are allowed.
For example, state $C = 3$ is disallowed if the morphology of
the rung is $\sigma = 2,3,4$, the state $C = 6$ is disallowed if the morphology of the rung 
OR of the rung 
immediately to the right is $\sigma = 3,4$, 
and so on.

We next construct the partition function of the track subject
to these constraints and also
the excluded volume constraints provided by the objects on the track.
Our transfer matrix formalism transfers
the state of a two-plaquette rung to the next two-plaquette rung to its {\it left}, subject to these constraints.
So, let $Z_n(C', \sigma')$ be the partition function of an $n$-rung track, where the leftmost rung 
is filled with the state $C'$, and has an underlying morphology $\sigma'$.  
Then, the partition function of the $(n+1)$-rung track, $Z_{n+1}(C, \sigma)$
is given by the recursion relation:
\begin{equation}
 Z_{n+1}(C, \sigma) = \sum_{C'} T_{\sigma, \sigma'}(C,C') Z_{n}(C',\sigma'),
\end{equation}
where  $T_{\sigma, \sigma'}(C,C')$ is a $6 \times 6$ transfer matrix, consistent with the excluded volume constraints
of $C$ and $C'$ and also with the constraints provided by the underlying morphology $\sigma, \sigma'$.
We therefore have 16 possible transfer matrices, based on these indices $\sigma, \sigma'$.
However, we note that if there is a complete disruption in the track ($\sigma = 4$),
the partition function of the track
breaks up into a product over partition functions of open chains.
We deal with these cases separately since, as we show later, the computational cost is greatly reduced in this case.

\begin{figure}[h!]
{\bf 1,1.}
\includegraphics[width=.1\columnwidth,angle=0]{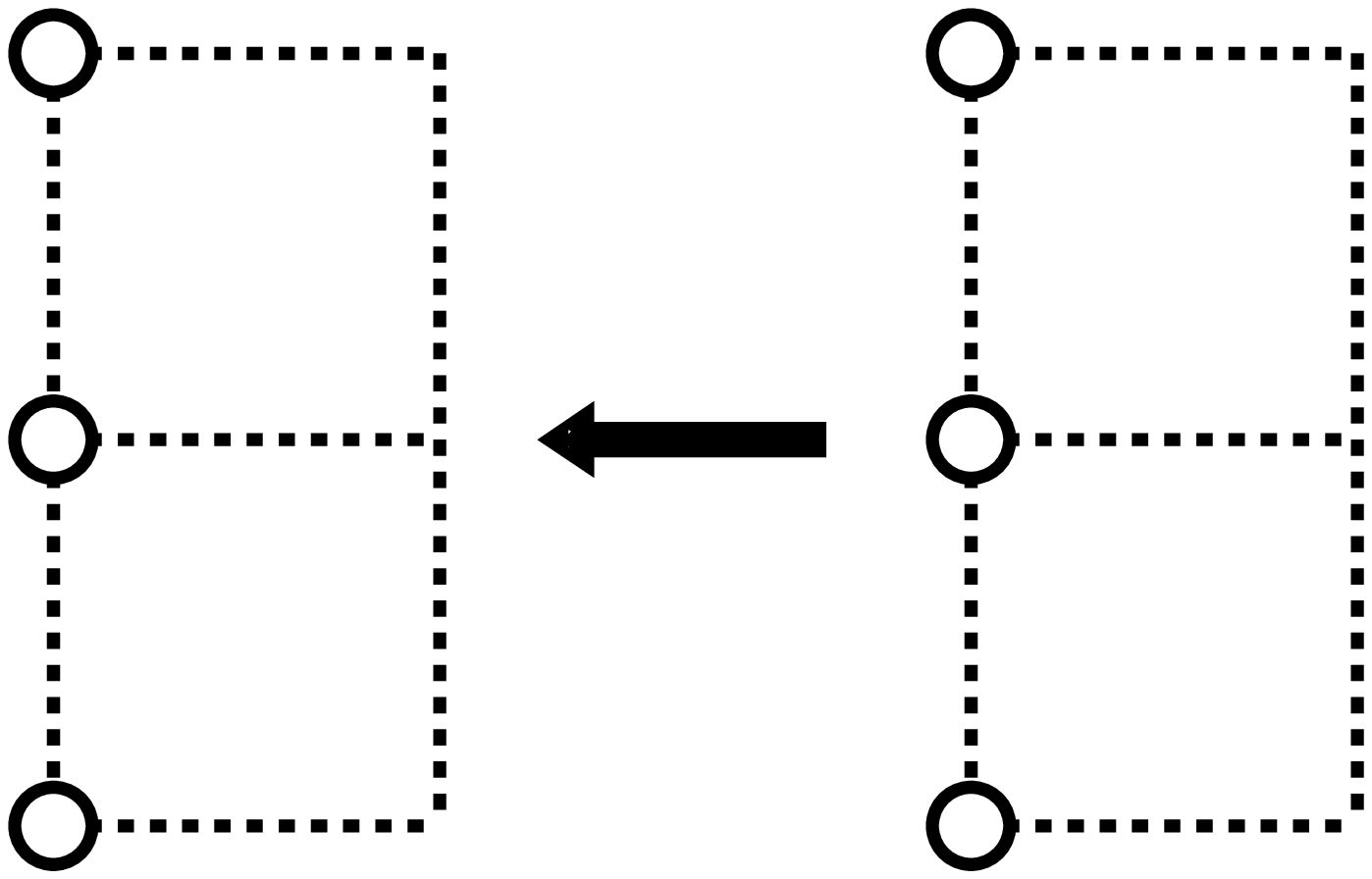}
\hspace{1cm}
{\bf 1,2.}
\includegraphics[width=.1\columnwidth,angle=0]{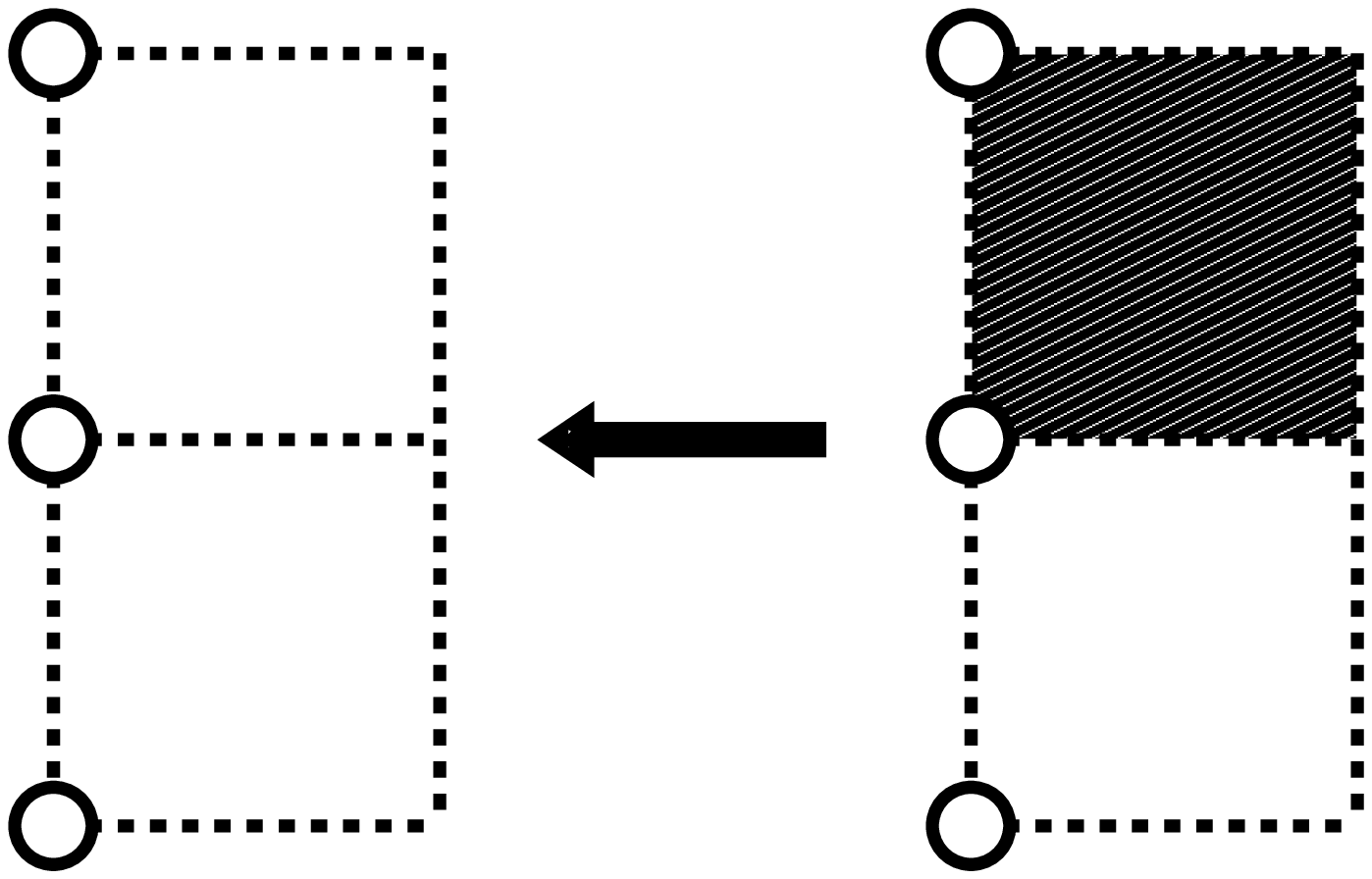}
\hspace{1cm}
{\bf 1,3.}
\includegraphics[width=.1\columnwidth,angle=0]{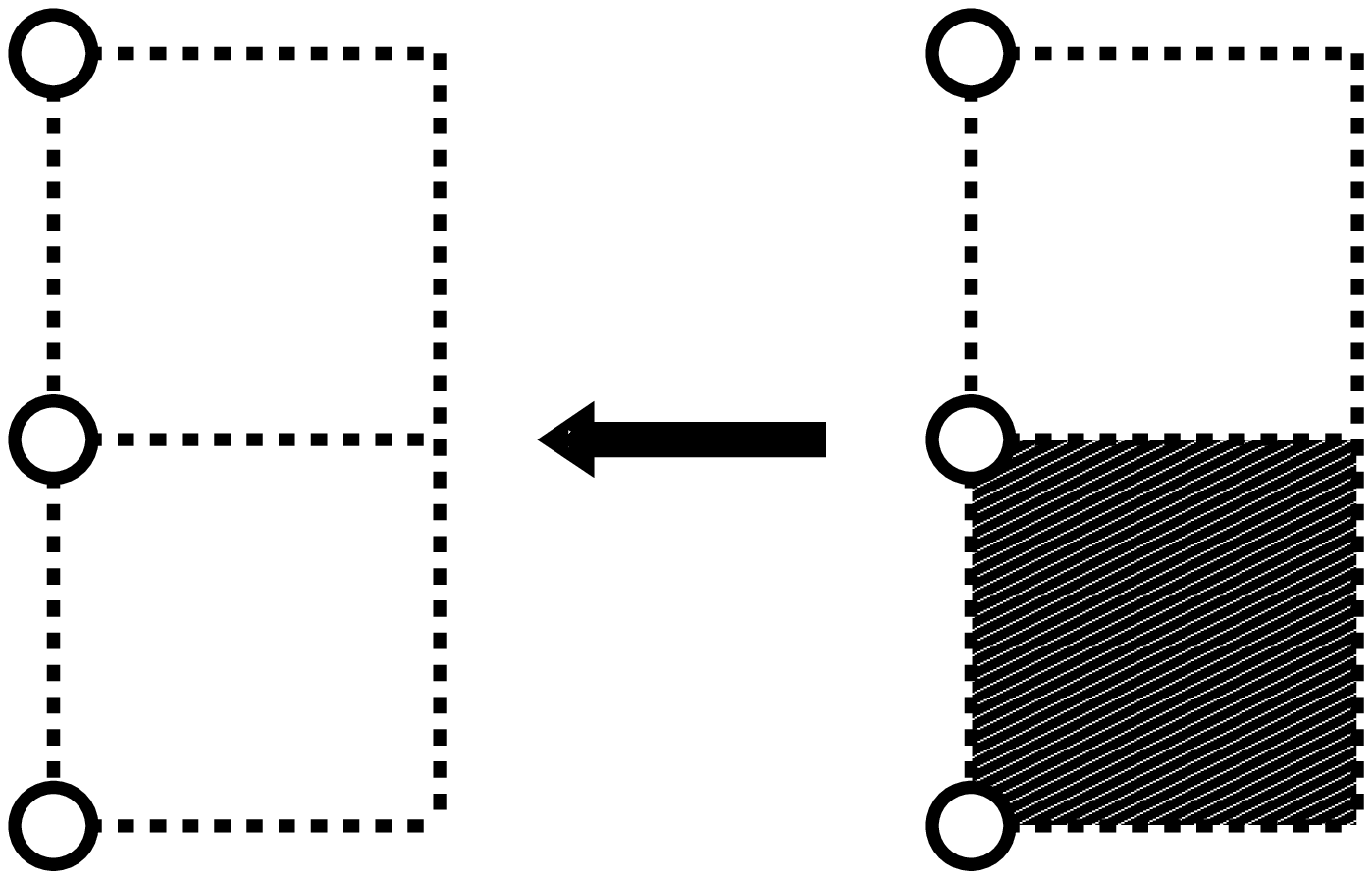}\\
\vspace{1cm}
{\bf 2,1.}
\includegraphics[width=.1\columnwidth,angle=0]{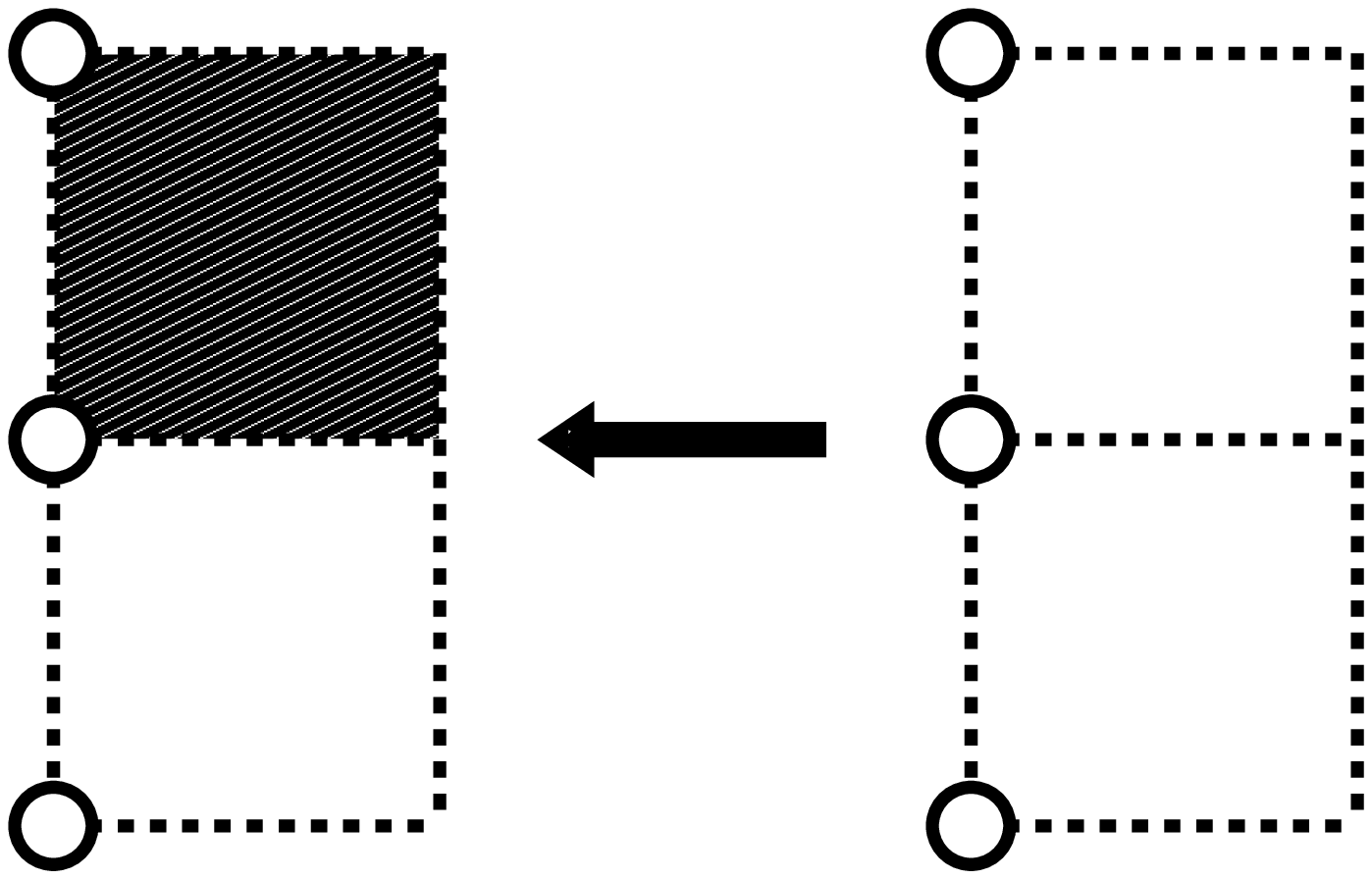}
\hspace{1cm}
{\bf 2,2.}
\includegraphics[width=.1\columnwidth,angle=0]{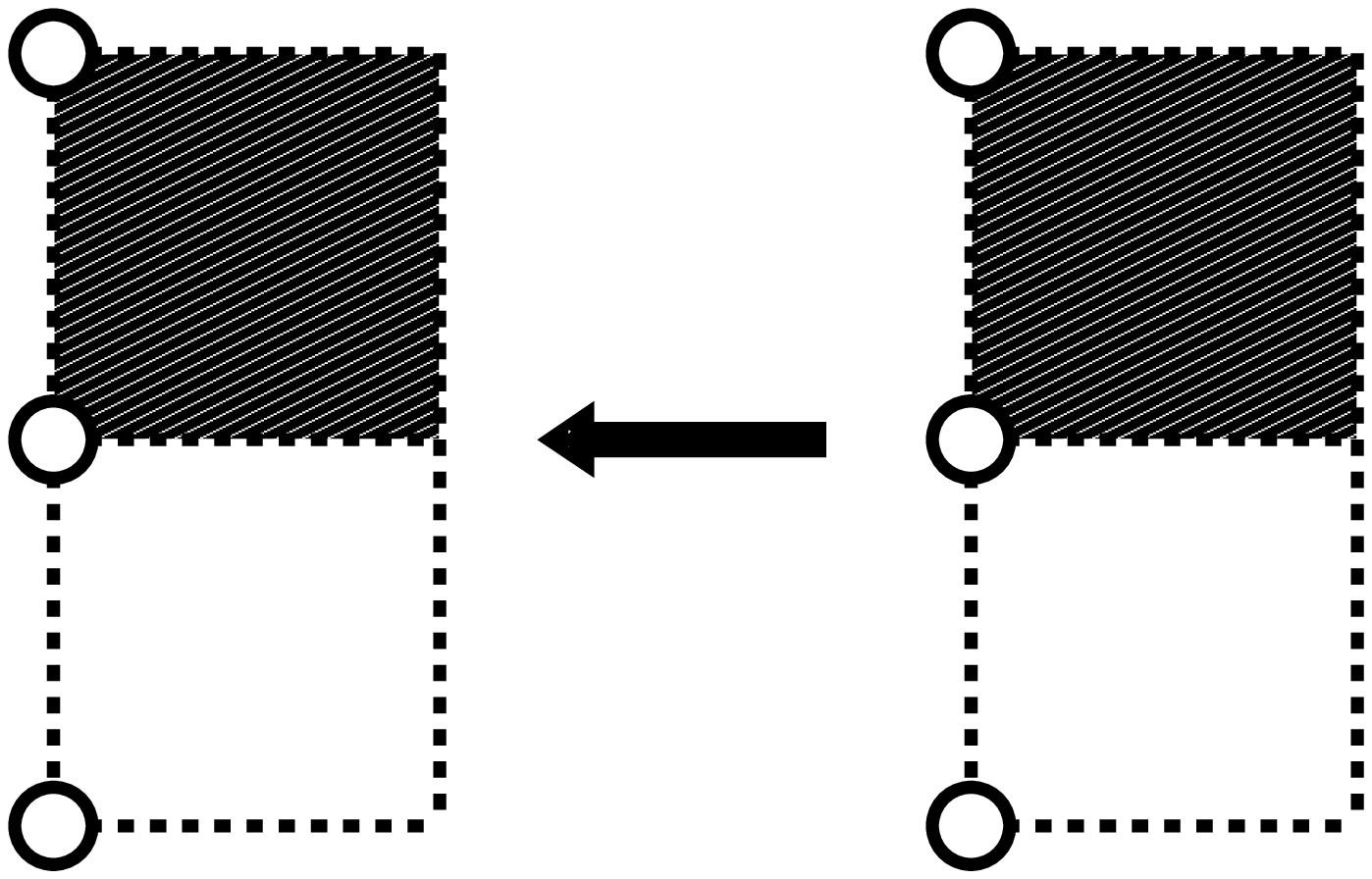}
\hspace{1cm}
{\bf 2,3.}
\includegraphics[width=.1\columnwidth,angle=0]{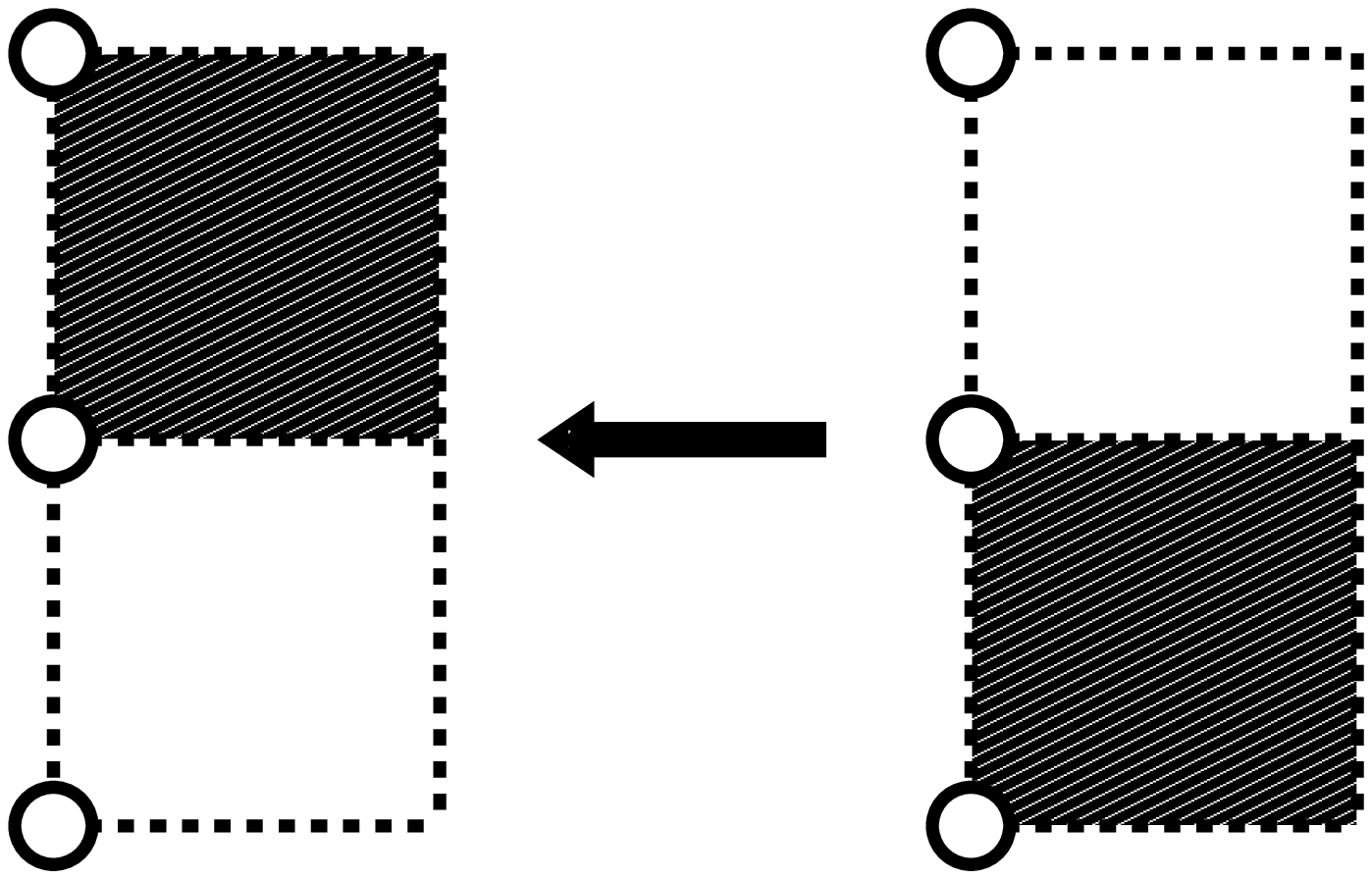}\\
\vspace{1cm}
{\bf 3,1.}
\includegraphics[width=.1\columnwidth,angle=0]{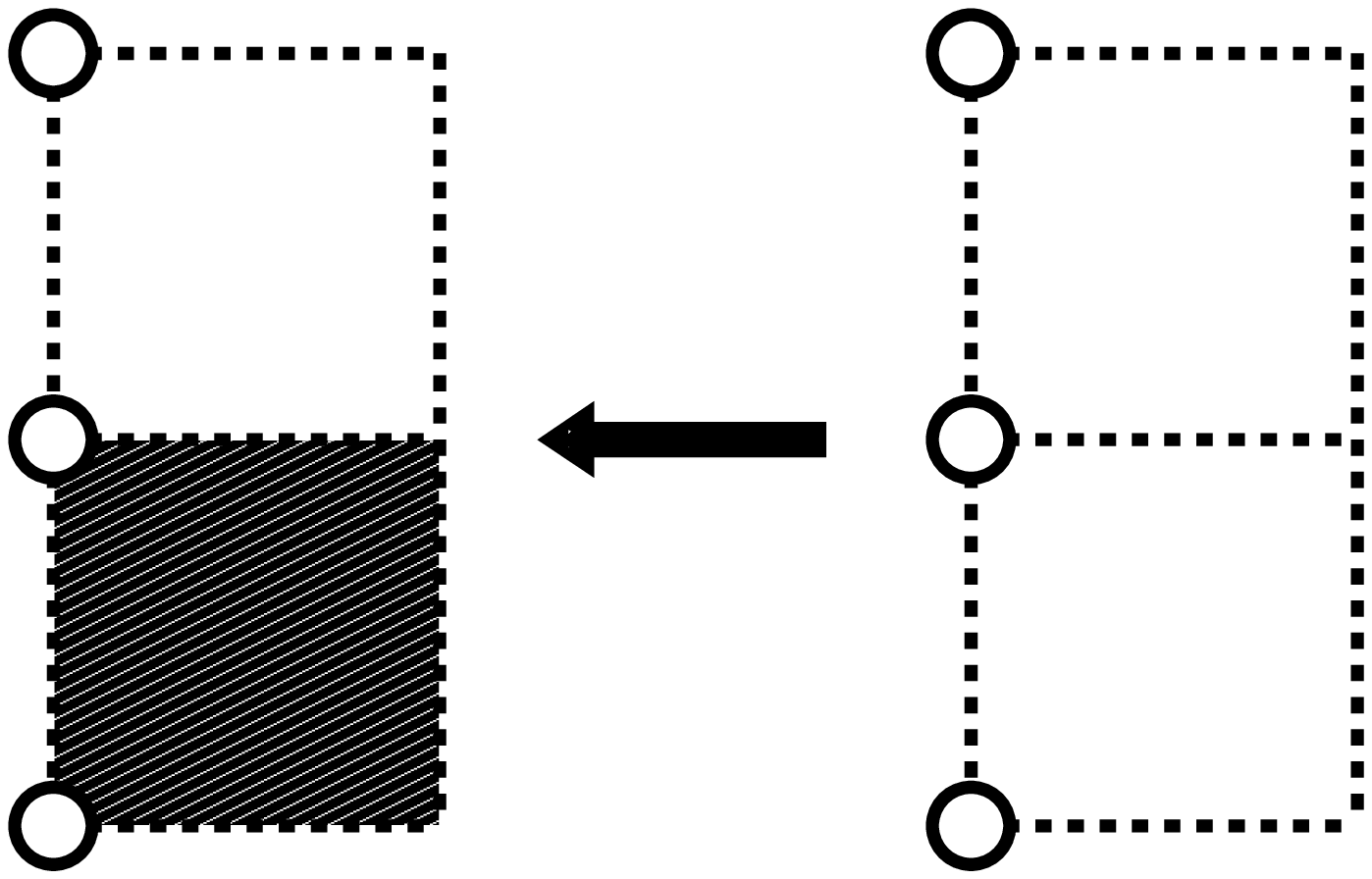}
\hspace{1cm}
{\bf 3,2.}
\includegraphics[width=.1\columnwidth,angle=0]{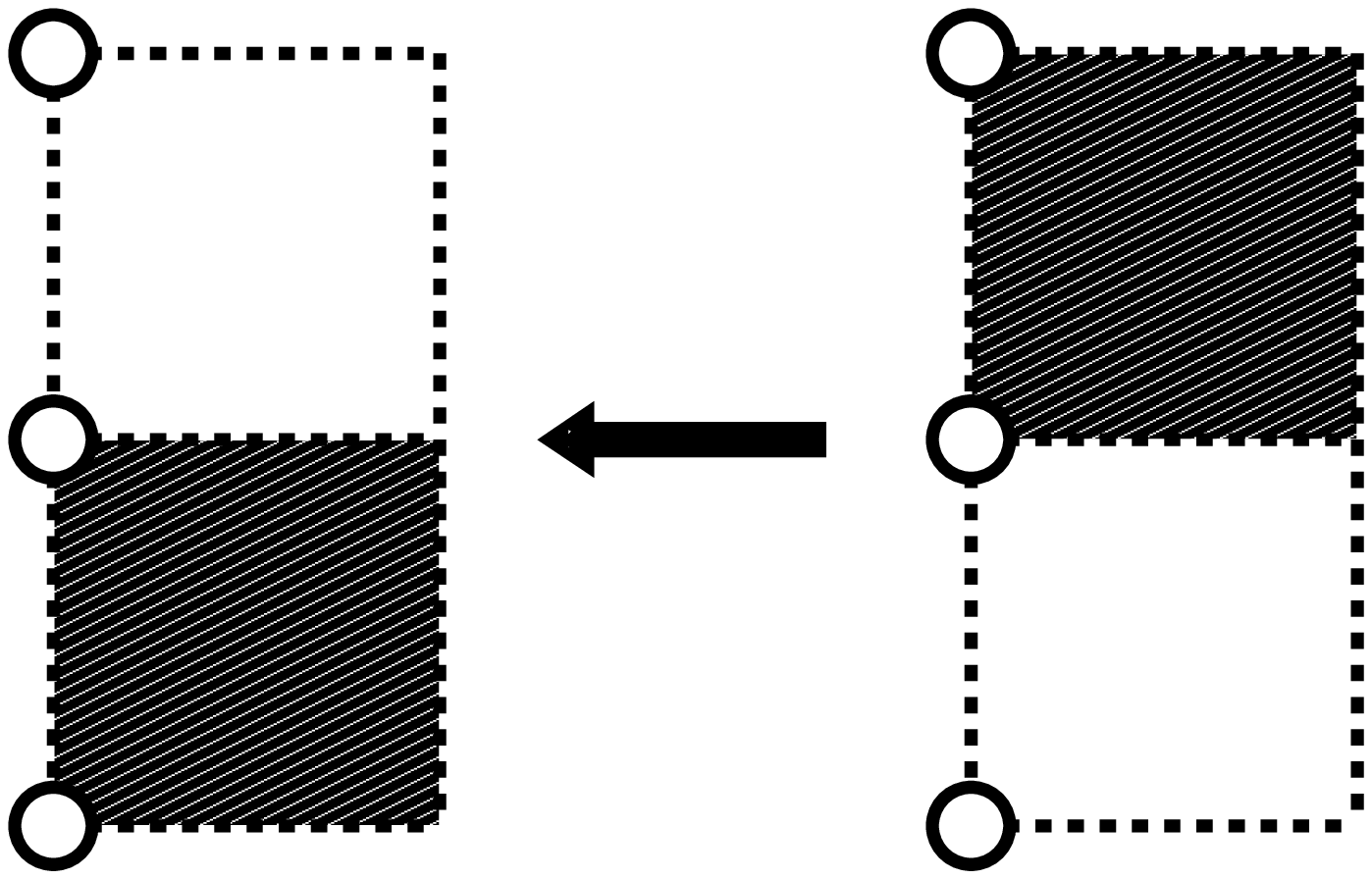}
\hspace{1cm}
{\bf 3,3.}
\includegraphics[width=.1\columnwidth,angle=0]{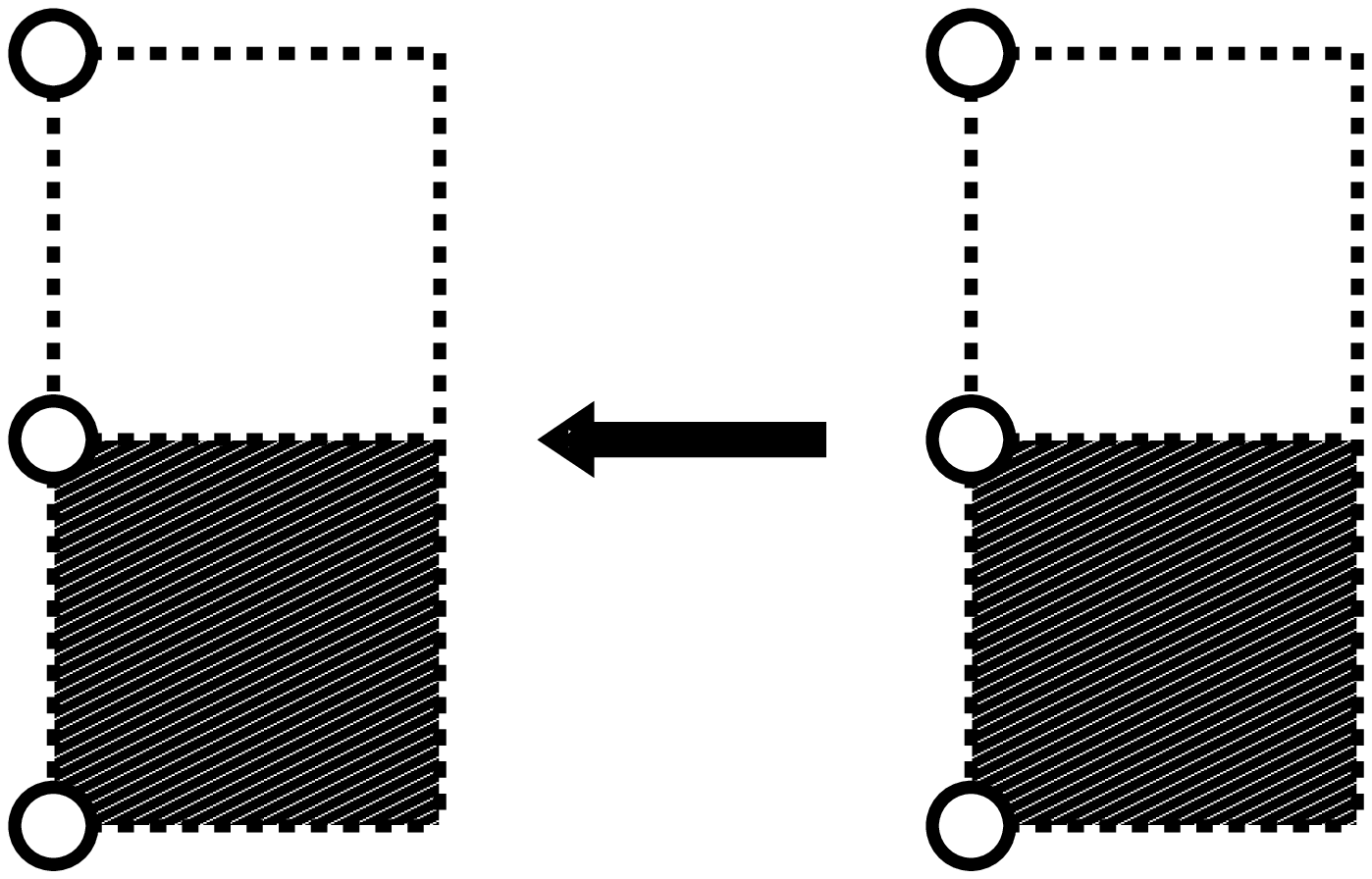}
\caption{The nine possible transfers between two rungs,
based on the combinations $\sigma, \sigma' = 1,2,3$.}
\label{transfer_possibilities_fig}
\end{figure}

Hence, based on the possible combinations $\sigma, \sigma' = 1,2,3$ (shown in Fig. \ref{transfer_possibilities_fig}),
there are {\bf nine} possible transfer matrices. We list them below:

\begin{equation}
\nonumber
\mathcal{T}_{1,1} = \left(
\begin{array}{cccccc}
 0 & 0 & 0 & z_s & 0 & 0 \\
 0 & 0 & 0 & z_d^2 & 0 & 0 \\
 z_d & z_d & z_d & z_d z_v^2 & z_d z_v & z_d z_v \\
 1 & 1 & 1 & z_v^2 & z_v & z_v \\
 0 & 0 & 0 & z_d z_v & 0 & z_d \\
 0 & 0 & 0 & z_d z_v & z_d & 0 \\
\end{array}
\right), 
\hspace{1cm}
\mathcal{T}_{1,2} = 
\left(
\begin{array}{cccccc}
 0 & 0 & 0 & 0 & 0 & 0 \\
 0 & 0 & 0 & 0 & 0 & 0 \\
 0 & 0 & 0 & z_d z_v & 0 & z_d \\
 0 & 0 & 0 & z_v & 0 & 1 \\
 0 & 0 & 0 & 0 & 0 & 0 \\
 0 & 0 & 0 & z_d & 0 & 0 \\
\end{array}
\right),
\hspace{1cm}
\mathcal{T}_{1,3} = 
\left(
\begin{array}{cccccc}
 0 & 0 & 0 & 0 & 0 & 0 \\
 0 & 0 & 0 & 0 & 0 & 0 \\
 0 & 0 & 0 & z_d z_v & z_d & 0 \\
 0 & 0 & 0 & z_v & 1 & 0 \\
 0 & 0 & 0 & z_d & 0 & 0 \\
 0 & 0 & 0 & 0 & 0 & 0 \\
\end{array}
\right),
\end{equation}

\begin{equation}
\nonumber
\mathcal{T}_{2,1} = 
\left(
\begin{array}{cccccc}
 0 & 0 & 0 & 0 & 0 & 0 \\
 0 & 0 & 0 & 0 & 0 & 0 \\
 0 & 0 & 0 & 0 & 0 & 0 \\
 1 & 1 & 1 & z_v^2 & z_v & z_v \\
 0 & 0 & 0 & 0 & 0 & 0 \\
 0 & 0 & 0 & z_d z_v & z_d & 0 \\
\end{array}
\right),
\hspace{1cm}
\mathcal{T}_{2,2} = 
\left(
\begin{array}{cccccc}
 0 & 0 & 0 & 0 & 0 & 0 \\
 0 & 0 & 0 & 0 & 0 & 0 \\
 0 & 0 & 0 & 0 & 0 & 0 \\
 0 & 0 & 0 & z_v & 0 & 1 \\
 0 & 0 & 0 & 0 & 0 & 0 \\
 0 & 0 & 0 & z_d & 0 & 0 \\
\end{array}
\right),
\hspace{1cm}
\mathcal{T}_{2,3} = 
\left(
\begin{array}{cccccc}
 0 & 0 & 0 & 0 & 0 & 0 \\
 0 & 0 & 0 & 0 & 0 & 0 \\
 0 & 0 & 0 & 0 & 0 & 0 \\
 0 & 0 & 0 & z_v & 1 & 0 \\
 0 & 0 & 0 & 0 & 0 & 0 \\
 0 & 0 & 0 & 0 & 0 & 0 \\
\end{array}
\right),
\end{equation}

\begin{equation}
\mathcal{T}_{3,1}=
\left(
\begin{array}{cccccc}
 0 & 0 & 0 & 0 & 0 & 0 \\
 0 & 0 & 0 & 0 & 0 & 0 \\
 0 & 0 & 0 & 0 & 0 & 0 \\
 1 & 1 & 1 & z_v^2 & z_v & z_v \\
 0 & 0 & 0 & z_d z_v & 0 & z_d \\
 0 & 0 & 0 & 0 & 0 & 0 \\
\end{array}
\right),
\hspace{1cm}
\mathcal{T}_{3,2} = 
\left(
\begin{array}{cccccc}
 0 & 0 & 0 & 0 & 0 & 0 \\
 0 & 0 & 0 & 0 & 0 & 0 \\
 0 & 0 & 0 & 0 & 0 & 0 \\
 0 & 0 & 0 & z_v & 0 & 1 \\
 0 & 0 & 0 & 0 & 0 & 0 \\
 0 & 0 & 0 & 0 & 0 & 0 \\
\end{array}
\right),
\hspace{1cm}
\mathcal{T}_{3,3} = 
\left(
\begin{array}{cccccc}
 0 & 0 & 0 & 0 & 0 & 0 \\
 0 & 0 & 0 & 0 & 0 & 0 \\
 0 & 0 & 0 & 0 & 0 & 0 \\
 0 & 0 & 0 & z_v & 1 & 0 \\
 0 & 0 & 0 & z_d & 0 & 0 \\
 0 & 0 & 0 & 0 & 0 & 0 \\
\end{array}
\right).
\end{equation}
It is useful to note that our convention of assigning objects to the rungs, leads to factors of $z_v$ appearing
asymmetrically in various entries of the transfer matrices. For example $\mathcal{T}_{1,1}(4,5) = z_v$,
since this leaves a vacancy at the bottom of the rung transfered from.

\subsubsection{Partition Function}

If there are no complete blockades on the track,
the partition function is that of a closed chain given by
\begin{equation}
Z^{closed}_{\textmd{track}} = \textmd{Tr}(\mathcal{T}_L ....... \mathcal{T}_3 \mathcal{T}_2 \mathcal{T}_1). 
\end{equation}
where $L$ is the size of the lattice and the matrices $\mathcal{T}_i$ are chosen according to the underlying morphology as described above.
Here $\mathcal{T}_1 = \mathcal{T}_{\sigma_2,\sigma_1}, \mathcal{T}_2 = \mathcal{T}_{\sigma_3,\sigma_2} ...,
\mathcal{T}_L = \mathcal{T}_{\sigma_{1},\sigma_{L}}$.

If one or more of the rungs on the track is completely blocked ($\sigma = 4$), then the partition function of the track
is given by a product of partition functions of open chains.
For an open chain where $N < L$ consecutive rungs are allowed for occupation,
the partition function is given by
\begin{equation}
Z^{open}_{\textmd{track}} = 
\langle \mathcal{L}_{\sigma_N} |\mathcal{T}_{N-1} ..... \mathcal{T}_3 \mathcal{T}_2 \mathcal{T}_1| \mathcal{R}_{\sigma_1} \rangle,
\end{equation}
where $\sigma_1$ and $\sigma_N$ represent the morphology of the first and $N$-th rung respectively. 
The three right vectors are given by (formally $\mathcal{R}_\sigma (C) = \mathcal{T}_{\sigma,4}(C,4)$)
\begin{equation}
 | \mathcal{R}_1 \rangle = 
 \left(
\begin{array}{c}
 0 \\
 0 \\
 z_d \\
 1 \\
 0 \\
 0 \\
\end{array}
\right),
\hspace{1cm}
| \mathcal{R}_2 \rangle = 
\left(
\begin{array}{c}
 0 \\
 0 \\
 0 \\
 1 \\
 0 \\ 
 0 \\
\end{array}
\right),
\hspace{1cm}
 | \mathcal{R}_3 \rangle = 
\left(
\begin{array}{c}
 0 \\
 0 \\
 0 \\
 1 \\
 0 \\
 0 \\
\end{array}
\right),
\end{equation}
and the three left vectors are given by (formally  $\mathcal{L}_\sigma(C) =\mathcal{T}_{4,\sigma}(4,C)$)
\begin{equation}
 \langle \mathcal{L}_1 | = 
\left(
\begin{array}{cccccc}
 1 & 1 & 1 & z_v^2 & z_v & z_v \\
\end{array}
\right),
\hspace{1cm}
\langle \mathcal{L}_2 | = 
\left(
\begin{array}{cccccc}
 0 & 0 & 0 & z_v & 0 & 1 \\
\end{array}
\right),
\hspace{1cm}
 \langle \mathcal{L}_3 | = 
\left(
\begin{array}{cccccc}
 0 & 0 & 0 & z_v & 1 & 0 \\
\end{array}
\right).
\end{equation}
\subsection{Choosing a New Configuration}
In order to choose a new configuration of objects on this track, we use the following recursive technique.
\subsubsection{Open Chain}
For an open chain, the state $C_N$ of the leftmost rung is chosen with the probability
\begin{equation}
p(C_N = i) = \frac{\langle \mathcal{L}_{\sigma_N}| i \rangle \langle i |\mathcal{T}_{N-1} ..... \mathcal{T}_3 \mathcal{T}_2 \mathcal{T}_1| \mathcal{R}_{\sigma_1} \rangle}
{\sum_{i} \langle \mathcal{L}_{\sigma_N}| i \rangle \langle i |\mathcal{T}_{N-1} ..... \mathcal{T}_3 \mathcal{T}_2 \mathcal{T}_1| \mathcal{R}_{\sigma_1} \rangle},
\end{equation}
where $| i \rangle$ are the standard $6 \times 1$ basis vectors.
Given this state $| i \rangle$ of the leftmost rung, the state $C_{N-1}$ of the next rung to the right, is then chosen with the probability
\begin{equation}
p(C_{N-1} = j) = \frac{ \langle \mathcal{L}' |j \rangle \langle j| \mathcal{T}_{N-2} ..... \mathcal{T}_3 \mathcal{T}_2 \mathcal{T}_1 |\mathcal{R}_{\sigma_1} \rangle}
{\sum_{j} \langle \mathcal{L}' |j \rangle \langle j| \mathcal{T}_{N-2} ..... \mathcal{T}_3 \mathcal{T}_2 \mathcal{T}_1 | \mathcal{R}_{\sigma_1} \rangle},
\end{equation}
where $\langle \mathcal{L}'| = \langle i |\mathcal{T}_{N-1}$, acts as the new left vector.
We can then recursively populate the entire track using this procedure.
Clearly, starting from a given right vector $|\mathcal{R}_{\sigma_1} \rangle$ depending on the morphology of the rightmost
rung, one only needs to
store the partial products $\mathcal{T}_{k} ..... \mathcal{T}_3 \mathcal{T}_2 \mathcal{T}_1|\mathcal{R}_{\sigma_1} \rangle$,
of $6 \times 1$ vectors at each rung in this algorithm.
\subsubsection{Closed Chain}
For a closed chain, the state $C_L$ of the first rung is chosen with the probability
\begin{equation}
p(C_L = i) = \frac{\langle i |\mathcal{T}_{L} ..... \mathcal{T}_3 \mathcal{T}_2 \mathcal{T}_1| i \rangle}
{\sum_{i}\langle i |\mathcal{T}_{L} ..... \mathcal{T}_3 \mathcal{T}_2 \mathcal{T}_1| i \rangle}.
\end{equation}
Given this state  $|i \rangle$ of the first rung, the state $C_{L-1}$ of the next rung to the right is then chosen with the probability
\begin{equation}
p(C_{L-1} = j) = \frac{\langle i| \mathcal{T}_{L} |j \rangle \langle j| \mathcal{T}_{L-1} ..... \mathcal{T}_3 \mathcal{T}_2 \mathcal{T}_1 |i \rangle}
{\sum_{j} \langle i| \mathcal{T}_{L} |j \rangle \langle j| \mathcal{T}_{L-1} ..... \mathcal{T}_3 \mathcal{T}_2 \mathcal{T}_1 | i \rangle},
\end{equation}
and similarly for the rest of the chain (as for the open chain), until the entire track is filled. Thus, 
in the case of a closed chain, one needs to store the partial products of $6 \times 6$ matrices at each rung.
\begin{figure}
{\bf a.}
\includegraphics[width=.42\columnwidth,angle=0]{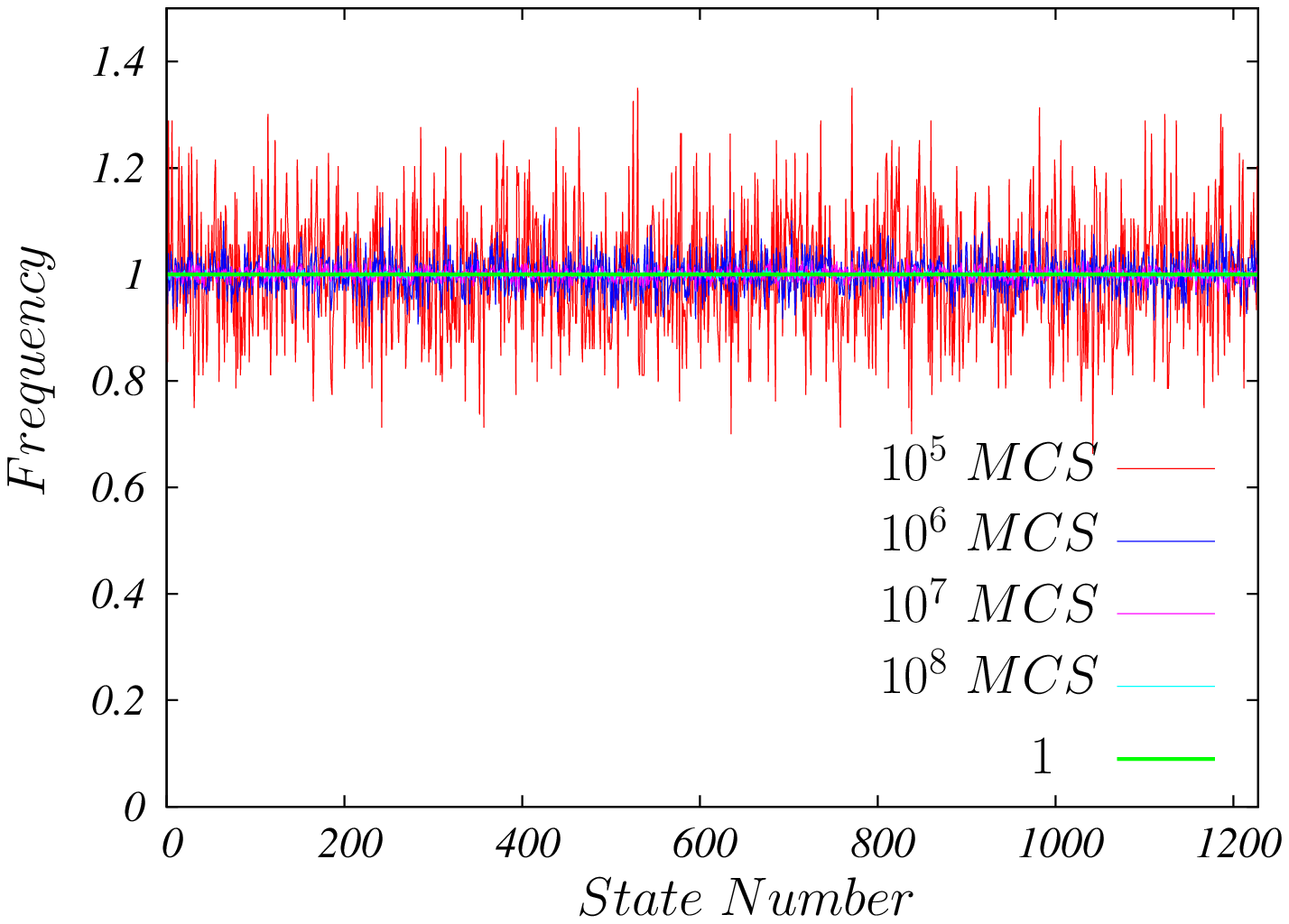}
{\bf b.}
\includegraphics[width=.42\columnwidth,angle=0]{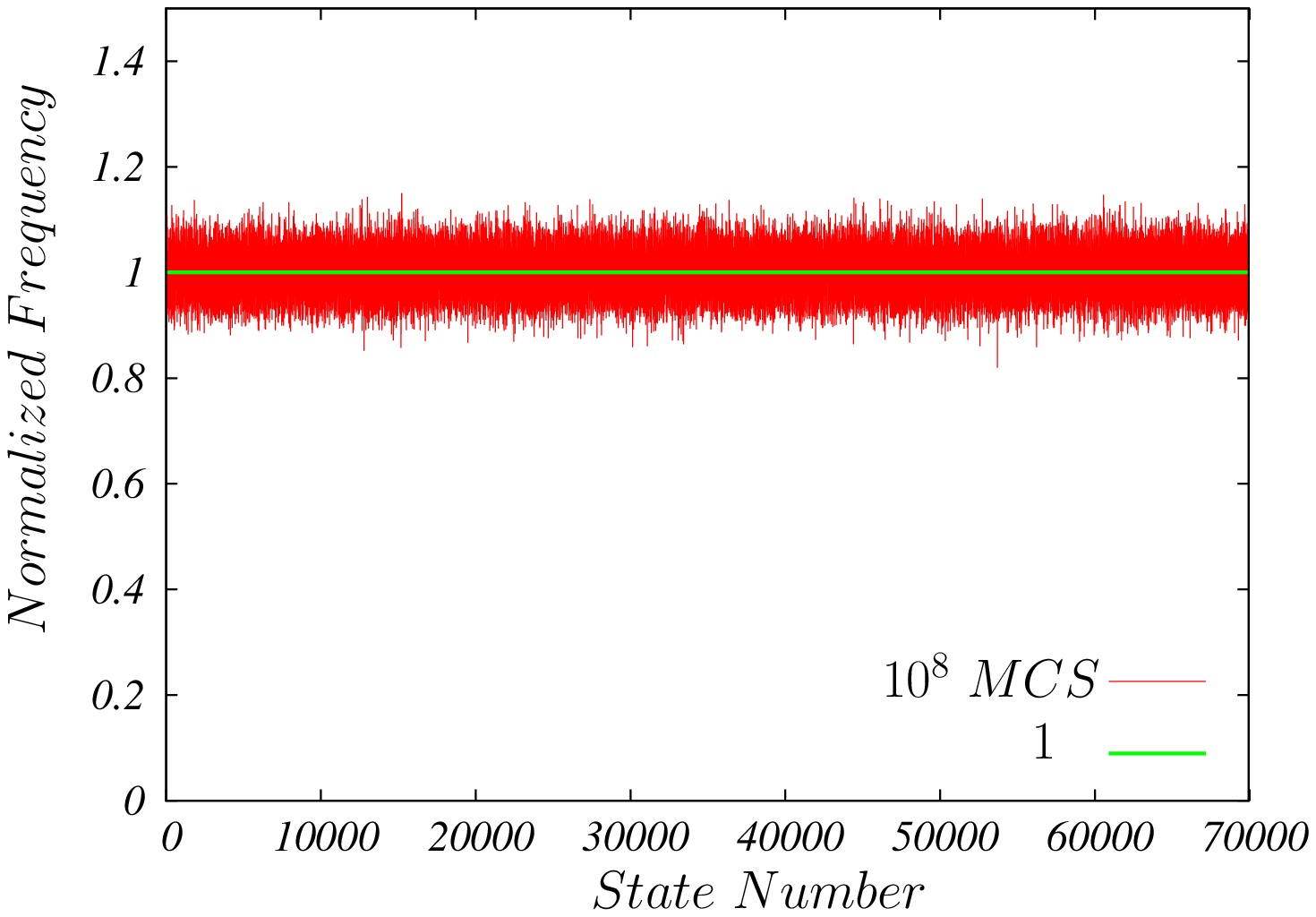}
\caption{Frequency of occurrence of states in a Monte Carlo simulation of a $4 \times 4$ periodic lattice {\bf a.} at full-packing
{\bf b.} for a general $v \ne 0$ (normalized by its weight in the partition function). All states are sampled with a frequency approaching $1$
as the number of Monte Carlo steps (MCS) are increased.}
\label{frequency_fig}
\end{figure} 

We note that the algorithm described above {\it does not reject} any configurations. 
This is particularly useful when studying high density phases, where local algorithms often encounter  ``jamming''.
This algorithm is naturally extendable to updates of wider tracks, where the size and the number of the transfer matrices grows
with the number of rows/columns considered. 
We also note that this algorithm is quite computationally efficient. For large lattice sizes and high densities,
the probability of encountering a periodic track falls rapidly. To update a single open track,
only storage of order $6 L$ numbers is required in an $L \times L$ system. In the
rare cases when we encounter a periodic track, we need storage of order $36L$ to
update it.  Naturally, the rarity of periodic tracks also implies that the algorithm
does not change winding sectors (defined exactly as in the usual dimer model) easily
for a large system. This is in principle a draw-back compared to loop algorithms or pocket-algorithms, 
both of which can be readily generalized for use in the present problem, and
may change sectors more easily (we have not explored this in any detail). 

In our simulations, we always start in the zero-winding sector, and our results for the larger values
of $L$ shown in the main text are therefore
averages over the zero-winding sector. However, as is well-known in
the context of interacting dimer models, the restriction to zero-winding in
the microscopic model simply corresponds to periodic boundary conditions for the coarse-grained heights. Therefore, it does not change our conclusions.
Finally, we note that a full Monte-Carlo sweep, requiring us to randomly choose $\mathcal {O}(L)$
different tracks and update their interior configurations, requires of order $\mathcal{O}(L^2)$
operations, making the time required comparable to that of other available schemes, while
being rejection-free.

\subsection{Detailed Balance and Ergodicity}
Since the new configurations are chosen with the correct weights from the ``restricted'' partition function,
this algorithm trivially satisfies the detailed balance criterion.
The question of ergodicity is more subtle. To check that the algorithm samples all available states of the system, 
we have performed the following numerical check.

We enumerate all possible states on a $4 \times 4$ lattice with periodic boundary conditions.
For the full-packing case (no vacancies), there are $1228$ possible configurations of squares and dimers. 
Using this explicit knowledge of all the states, we monitor the frequency with which each state is 
sampled in our simulations. We choose activities such that all fully-packed states
have unit Boltzmann weight and states with vacancies have zero weight.
In this case, we have checked that for a 
large enough number of Monte Carlo steps, all allowed states are sampled with
equal frequency. In addition we have checked that the variance of this frequency 
decreases as $\frac{1}{N_{MC}}$, 
where $N_{MC}$ represents the number of Monte Carlo steps.
In Fig. \ref{frequency_fig} {\bf a}. we plot this frequency
table for different numbers of Monte Carlo samplings.

We have also enumerated all possible states for this small sample when $v \ne 0$. In this case  there are $69941$ configurations of dimers, squares and vacancies available to the system. We check explicitly that each one of these states is sampled with the correct probability given by
\begin{equation}
p(\mathcal{C}^{*}_{dsv}) = \frac{w^{N^{*}_d} v^{N^{*}_v}}{\sum_{\mathcal{C}_{dsv}} w^{N_d} v^{N_v}},
\end{equation}
where $N^{*}_d$ and $N^{*}_v$ are the number of dimers and vacancies in the configuration $\mathcal{C}^{*}_{dsv}$,
and $N_d$ and $N_v$ are the number of dimers and vacancies in the configuration $\mathcal{C}_{dsv}$.
The sum is over all possible
configurations of the system.
In Fig. \ref{frequency_fig} {\bf b}., we plot the frequency of the occurrence of each of the $69941$ configurations
in our simulations, normalized by the above probability. We find that the normalized frequency of each of
these states converges to $1$, confirming the ergodicity of our algorithm (for small lattice sizes).

\section{Additional Numerical Evidence}

Finally, we use our Monte Carlo update scheme to perform large scale simulations
on the lattice gas of dimers and squares on the square lattice.
Recent simulations of the hard-square lattice gas have shown the necessity
of simulations on large system sizes to fully understand the nature of scaling in such
hard-core systems with columnar ordering
\cite{Feng_Blote_Nienhuis, thesis}.
The columnar ordered state is relatively unstable to the presence of vacancies, as compared to sublattice ordering,
and is characterised by large correlation lengths.
We therefore perform simulations on lattices of sizes up to $1024 \times 1024$ in order to
fully elucidate the phase diagram of this
system. 

\begin{figure}[h!]
{\includegraphics[width=.45 \columnwidth]{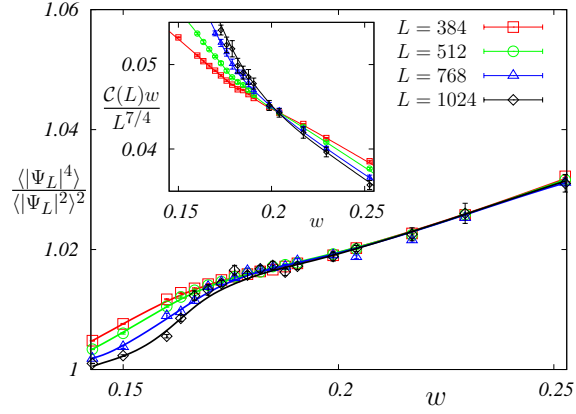}}
\caption{Phase transition along $I$, the fully-packed boundary $SD$, ($v=0$) corresponding to the pure squares and dimers mixture. The above figure shows the Binder-ratio $\langle |\Psi_L|^4\rangle/\langle |\Psi_L|^2\rangle^2$ ($\Psi_L \equiv \sum_{\vec{r} }\psi(\vec{r})$) sticking for $w >w_c^{(0)} \approx 0.198(2)$ signalling a $v=0$ power-law columnar ordered phase for $w>w_c^{(0)}$. Inset shows $\mathcal{C}(L) = \langle |\sum_{\vec{r}} \psi(\vec{r})|^2\rangle/L^2$ scaled by $L^{7/4}/w$ for various $L$. The curves cross at $w_c^{(0)}$, consistent with $\eta(w_c^{(0)}) = 1/4$.}
\label{Fullpacking_fig}
\end{figure}

\begin{figure}[h!]
{\bf (a)}
\includegraphics[width=.45\columnwidth,angle=0]{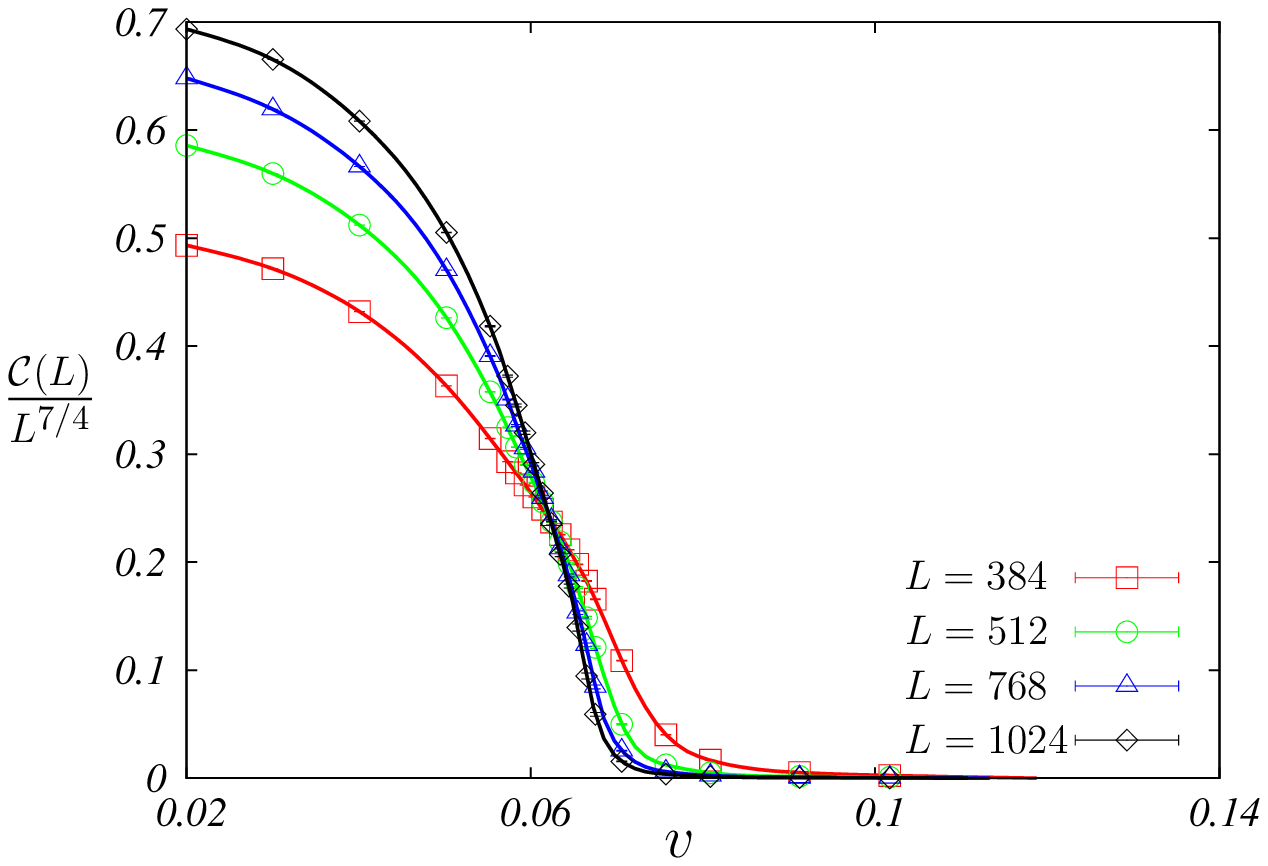}\\
{\bf (b)}
\includegraphics[width=.43\columnwidth,angle=0]{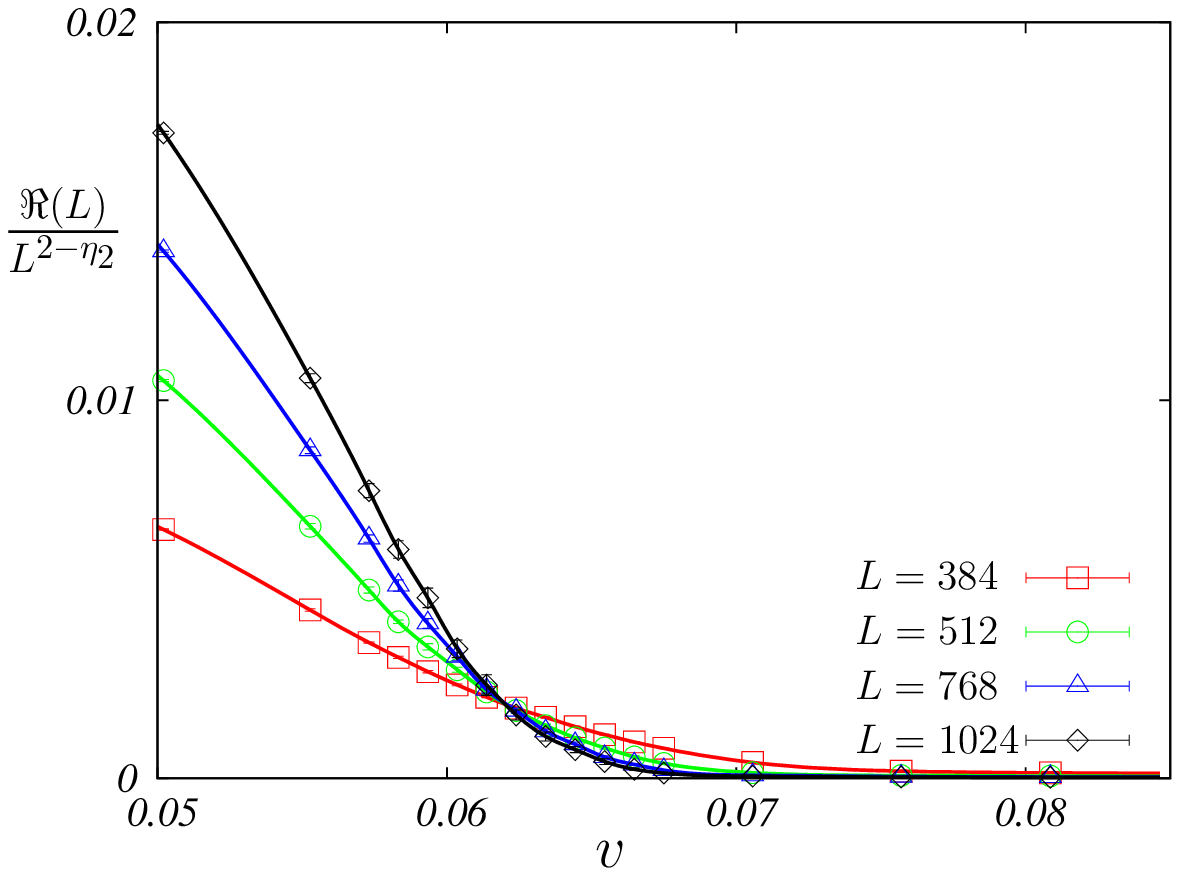}\\
{\bf (c)}
\includegraphics[width=.45\columnwidth]{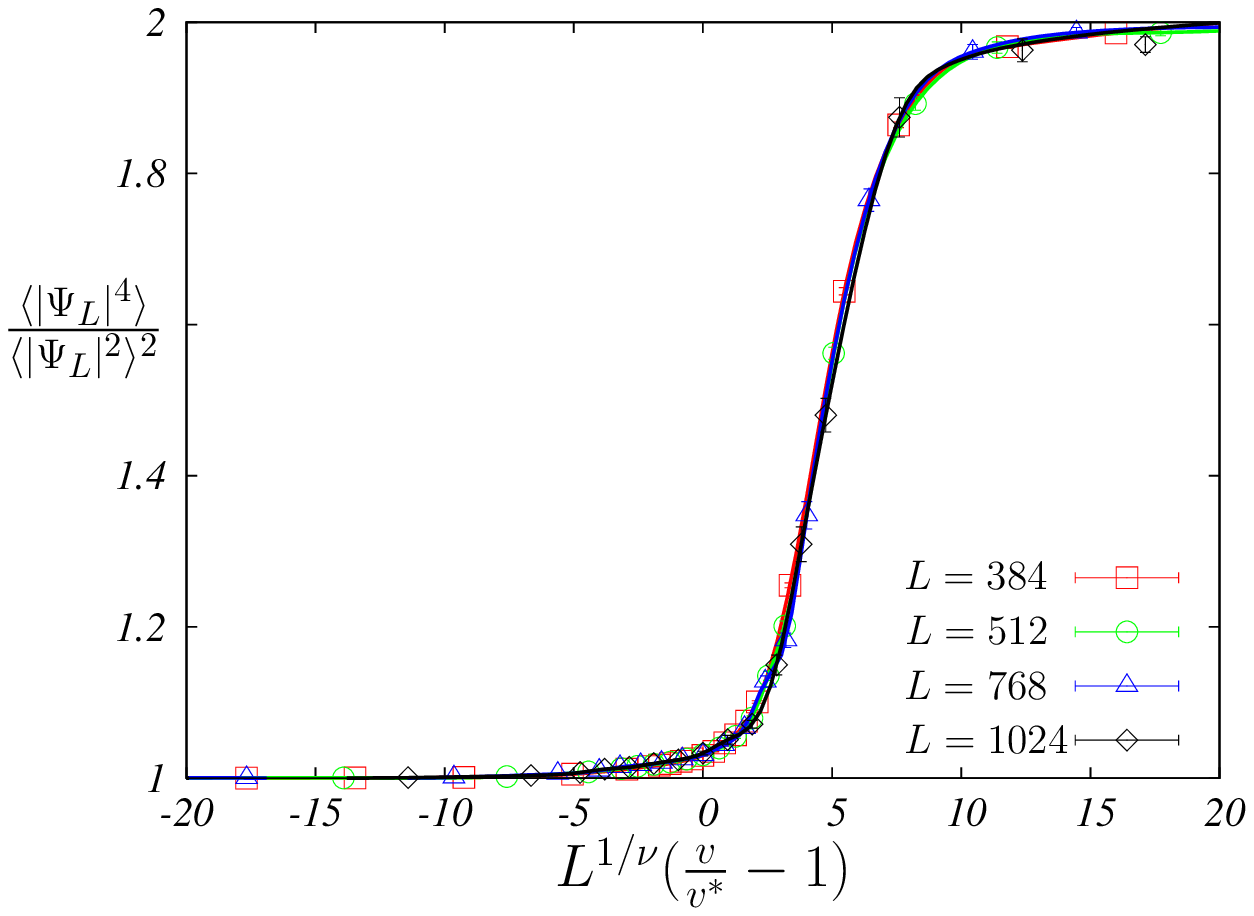}
\caption{Phase transition along $II$, defined in Eq. \ref{trajectory_eq}.
{\bf (a)} $\mathcal{C}(L) = \langle |\sum_{\vec{r}} \psi(\vec{r})|^2\rangle/L^2$ scaled by $L^{7/4}$ plotted as a function of $v$, 
for different values of the system size $L$. 
The curves show a sharp crossing, allowing us to estimate the location
of the critical point at $v_c = 0.0623(1)$ (the corresponding value
of $w_c$ is therefore $w_c = 0.1600(1)$).
{\bf (b)} $\Re(L) = \langle [\sum_{\vec{r}}{\rm Re}(\psi^2(\vec{r})) ]^2\rangle/L^2$ scaled by $L^{2-\eta_2}$  as a function of $v$
for different values of the system size $L$.
The curves again cross at the value of $v_c$ estimated above when $\eta_2$
is chosen as $\eta_2 = 0.70(5)$. 
(Inset) Scaling collapse of  $\Re(L)/L^{2-\eta_2}$ 
using the value $\nu = 1.70$ for the correlation length exponent. Note that these estimates of $\eta_2$ and $\nu$ satisfy the Ashkin-Teller relation $\eta_2 = 1-1/(2\nu)$ within errors.
{\bf (c)} Scaling collapse  with $\nu = 1.70(5)$ of Binder-ratio $\langle |\Psi_L|^4\rangle/\langle |\Psi_L|^2\rangle^2$ ($\Psi_L \equiv \sum_{\vec{r} }\psi(\vec{r})$) for various $L$.}
\label{Intermediate_figs}
\end{figure}

We use the convention $v = z_v/ \sqrt{z_s}, w = z_d/{z_s}^{1/4}$ and $z_s + z_d^2 + z_v^4 = 1$.
In our simulations, we focus on three cuts through the phase diagram (Fig. 2 of main text)
enumerated below.

\begin{figure}
{\bf (a)}
\includegraphics[width=.45\columnwidth,angle=0]{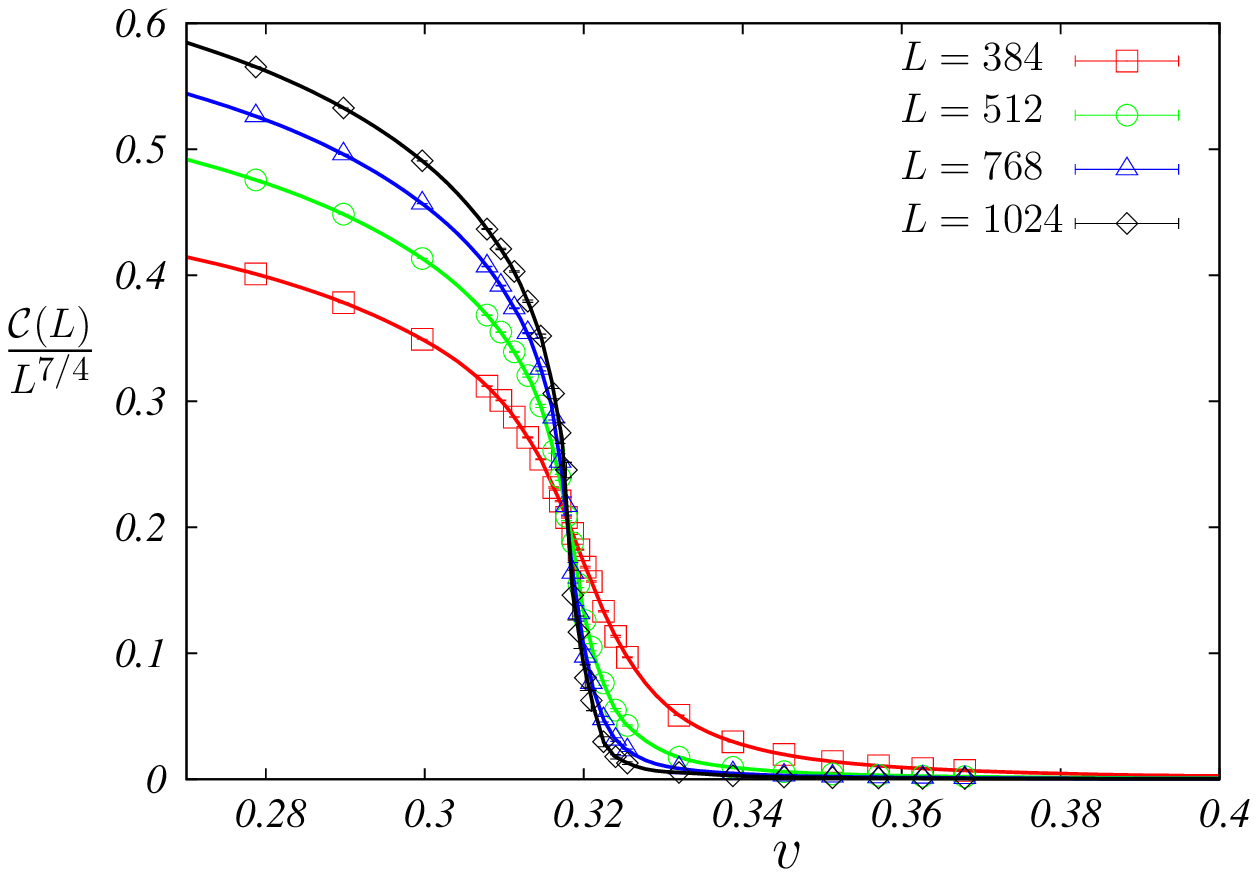}\\
{\bf (b)}
\includegraphics[width=.45\columnwidth,angle=0]{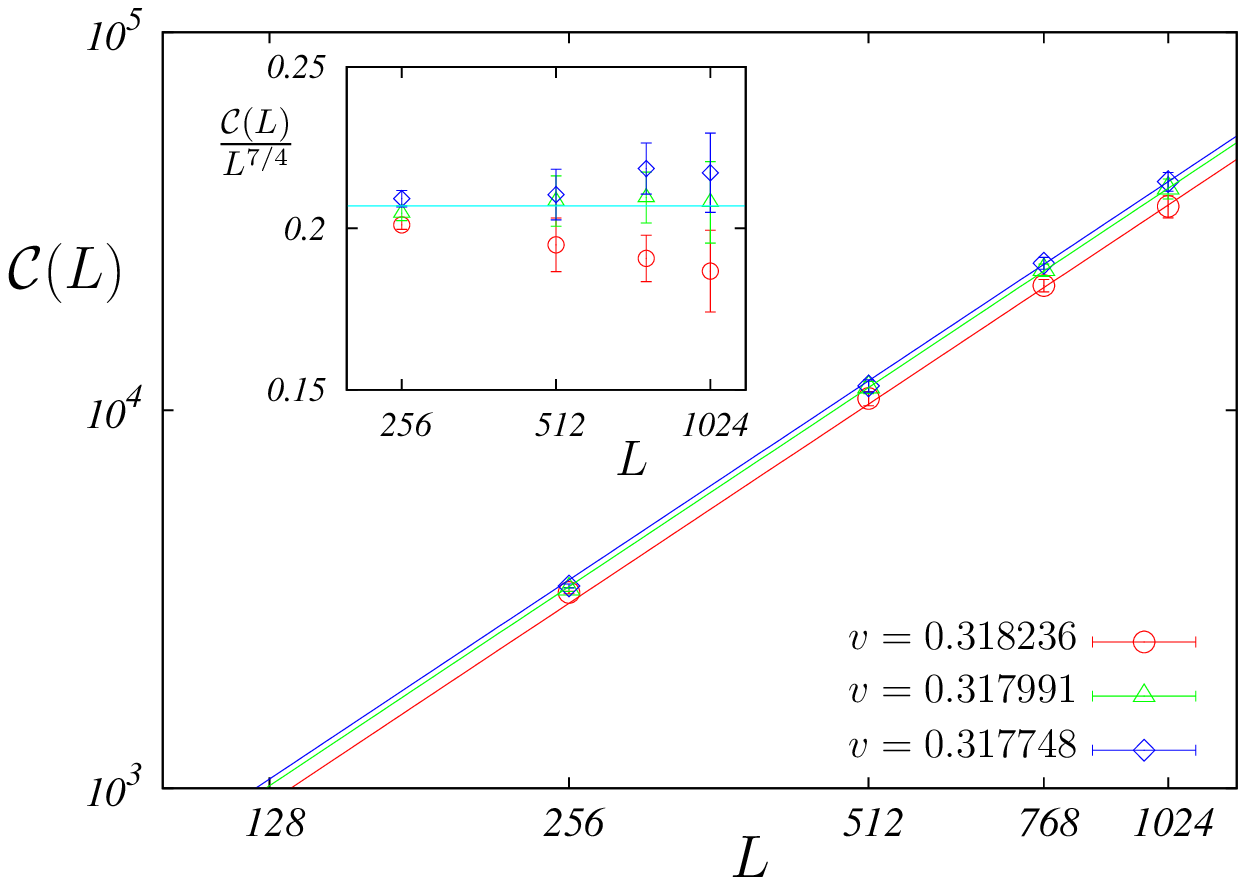}\\
{\bf (c)}
\includegraphics[width=.45\columnwidth,angle=0]{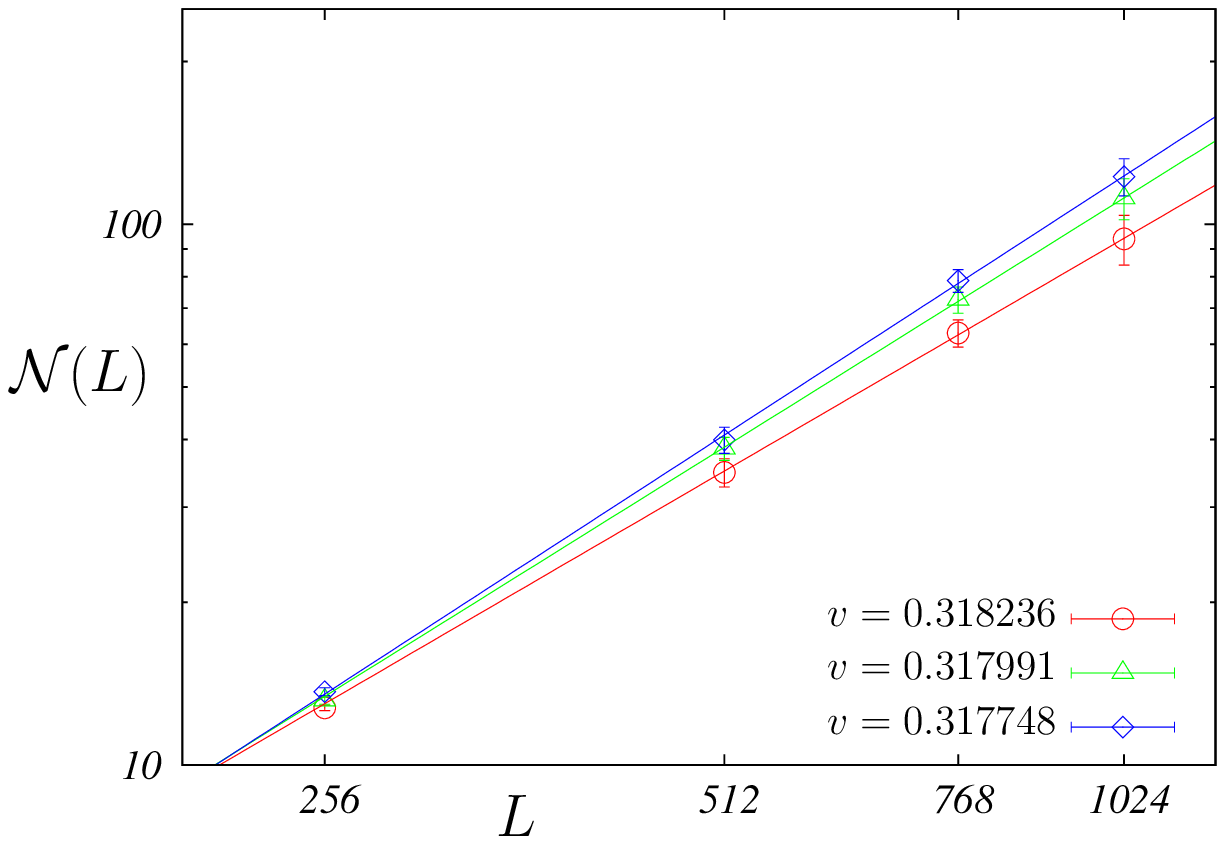}
\caption{Phase transition along $III$ (the boundary $VS$ with $w = 0$), corresponding to the hard-square lattice gas.
{\bf (a)}
The figure shows the columnar order parameter $\mathcal{C}(L) = \langle |\sum_{\vec{r}} \psi(\vec{r})|^2\rangle/L^2$ scaled by $L^{7/4}$ plotted as a function of $v$ for different values of the system size $L$.
The sharp crossing seen is consistent with the fact that $\eta = \frac{1}{4}$ at this
transition.
Our estimate of the transition point is $v_c^{*} = 0.3180(3)$ (see data in next figure).
{\bf (b)}
A precise estimate of the phase transition point $v_c^{*}$ along $III$ (the boundary $VS$ with $w = 0$), is obtained by comparing the quality of power-law fits of $\mathcal{C}(L)$ to the form $aL^{7/4}$ for various $v$ in the critical region.  The figure shows
the data at three values of $v$ along with the best power-law fit curves. The inset
shows the $L$ dependence of $\frac{\mathcal{C}(L)}{L^{7/4}}$ at these three values.
From this we see that the power-law fit to $aL^{7/4}$ works best
at $v_c^{*} = 0.317991$ (this corresponds to $z_s=97.8$, $z_v=1$, $z_d=0$ in the original
parametrization of $Z_{dsv}$). The value quoted in the main text, {\em i.e.}
$v_c^{*} = 0.3180(3)$, rounds off this value to four decimal places and includes an error estimate that corresponds to the
spacing between the values of $v$ at which we have measured this $L$ dependence.
{\bf (c)}  The figure shows the $L$ dependence of $\mathcal{N}(L) =  \langle (\sum_{\vec{r}} T(\vec{r}))^2\rangle/L^2$ (with $T(\vec{r})$ defined in main text) at three values of $v$ in the critical region, along with the best-fit power-law curves
$bL^{2-\eta_2}$ at these values of $v$. 
The best-fit values of the power-law exponent $\eta_2$ depends sensitively on the value of $v$ in this critical region. 
We find that the postulated power-law fit works
appreciably better at $v_c^{*} =  0.317991$ (identified in previous
figure the power-law form
$aL^{7/4}$) as compared
to neighbouring values of $v$. The corresponding
best-fit estimate of $\eta_2$ is $\eta_2^{*} = 0.46$.}
\label{hardsquare_fig}
\end{figure}

{\bf {\em I}:} The fully-packed boundary $SD$, ($v=0$) corresponding to the pure squares and dimers mixture.
 We find that the system exhibits a KT transition from the square-rich columnar ordered phase to a power-law
 ordered dimer-rich phase above the critical point $w_c^{(0)} = 0.198(2)$. The details of the power law correlations are provided 
 in Fig. 2 of the main text. In Fig. \ref{Fullpacking_fig} we display the sticking of the Binder ratio  $\langle |\Psi_L|^4\rangle/\langle |\Psi_L|^2\rangle^2$ ($\Psi_L \equiv \sum_{\vec{r} }\psi(\vec{r})$) for $w > w_c^{(0)}$
 along $I$, signalling a power-law ordered phase in this region.

{\bf {\em II}:} A trajectory passing through a generic point on the phase boundary separating the square-rich
columnar ordered phase from the disordered squares-dimers-vacancy fluid phase. In our simulations, we move
along the trajectory
\begin{equation}
z_d = \alpha z_v,
\label{trajectory_eq}
\end{equation}
where $\alpha \approx 2.54947$. This corresponds to the trajectory $w = \alpha v (1 + w^2 + v^4)^{1/4}$.
We find that in this case the transition is of second order, with a critical point at $P \equiv (w_c,v_c) = (0.1600(1),0.0623(1))$. 
In Fig. \ref{Intermediate_figs} {\bf (a)},
we display the critical crossing of the columnar order parameter $\mathcal{C}(L) = \langle |\sum_{\vec{r}} \psi(\vec{r})|^2\rangle/L^2$ scaled by $L^{2-\eta_2}$ at this critical point $P$,
consistent with Ashkin-Teller behaviour with $\eta = \frac{1}{4}$.
We find a good collapse of these curves with the scaling exponent $\nu = 1.70(5)$ (displayed in Fig. 4 of main text).
In Fig. \ref{Intermediate_figs} {\bf (b)} we display the critical crossing of the real part of the order parameter
$\Re(L) = \langle [\sum_{\vec{r}}{\rm Re}(\psi^2(\vec{r})) ]^2\rangle/L^2$ scaled by $L^{2-\eta_2}$,
with $\eta_2 = 0.70(5)$, at this point $P$ . Once again, these curves show a good collapse with the
scaling exponent $\nu = 1.70(5)$ (displayed in the inset of Fig. 4 of the main text). These estimates of $\eta_2$ and $\nu$
satisfy $2 \nu = (1-\eta_2)^{-1}$ within errors, as argued in the main text.
We also estimate $\nu = 1.70(5)$ from the scaling collapse of the Binder ratio (as displayed in Fig. \ref{Intermediate_figs} {\bf (c)}). 
The error estimates are
obtained by varying $\nu$ until the quality of the collapse deteriorates appreciably.

{\bf {\em III}:} The boundary $VS$, ($w =0$) corresponding to the hard-square lattice gas.
Once again, we find that the system displays a second order phase transition to a columnar ordered state as the density of squares is increased.
We display numerical results in the vicinity of this transition in Fig. \ref{hardsquare_fig}.
The first of these
figures shows the
columnar order parameter $\mathcal{C}(L) = \langle |\sum_{\vec{r}} \psi(\vec{r})|^2\rangle/L^2$ scaled by $L^{7/4}$ vs. $v$ for different values of the system size $L$. 
The curves display a sharp-crossing, which is consistent 
with $\eta = \frac{1}{4}$ at this transition. Using this as a prior, we obtain
the precise location of the critical point from the data displayed in the second figure.
This gives us the estimate $v_c^* = 0.3180(3)$. This is fed back into our analysis
of the data shown in the first figure, and used to collapse this data
into a scaling collapse with $\nu$ as an adjustible parameter (displayed in Fig. 3 of main text). This yields
the estimate $\nu^* = 0.92(3)$ quoted in the main text. 
The error-estimate is obtained
by varying $\nu$ until the data collapse deteriorates appreciably.
Finally, in the third figure,
we show the $L$ dependence of the alternate two-fold order parameter $T$
defined in the main text for the hard-square lattice gas. Fitting the $L$ dependence of
$\mathcal{N}(L) =  \langle (\sum_{\vec{r}} T(\vec{r}))^2\rangle/L^2$ to a power-law form $bL^{2-\eta_2^{*}}$
yields the estimate $\eta_2^{*} = 0.46(3)$ quoted in the main text.
The error-bar
on $\eta_2^{*}$ is
relatively large  because of the sensitive dependence of the best-fit $\eta_2$ on
the estimated value of $v_c^{*}$.







\end{appendix}
\clearpage
\end{widetext}

\end{document}